\documentclass[journal,compsoc]{IEEEtran}
\usepackage{psfrag,amsmath,amsfonts,amssymb,amsthm}
\usepackage{cite}
\usepackage{graphicx,subfigure,psfrag}
\usepackage{algorithm}
\usepackage{booktabs}
\usepackage{multirow}
\usepackage{multicol}
\usepackage{colortbl}
\usepackage{color}
\usepackage{balance}
\usepackage{enumitem}
\usepackage{algpseudocode}
\usepackage{balance}
\usepackage{slashbox}

\def\thetabf{\boldsymbol \theta}

\def\pbf{{\bf p}}

\def\Ac{{\cal A}}

\def\Cc{{\cal C}}
\def\Dc{{\cal D}}

\def\Ic{{\cal I}}

\def\Kc{{\cal K}}

\def\Mc{{\cal M}}
\def\Nc{{\cal N}}

\def\Qc{{\cal Q}}
\def\Rc{{\cal R}}
\def\Sc{{\cal S}}
\def\Tc{{\cal T}}

\def\Vc{{\cal V}}
\def\Wc{{\cal W}}

\def\eg{{\it e.g.,\ \/}}

\def\ie{{\it i.e.,\ \/}}
\def\nn{\nonumber}

\DeclareMathOperator*{\argmin}{argmin}

\theoremstyle{definition}

\newtheorem{Remark}{Remark}
\newtheorem{proposition}{Proposition}

\usepackage{bm}
\begin{document}

\title{Design and Optimization of Heterogeneous Coded Distributed Computing with \\Nonuniform File Popularity}

 \author{Yong~Deng, \IEEEmembership{Member, IEEE}  and Min~Dong, \IEEEmembership{Fellow, IEEE}
\thanks{Preliminary results in this work were presented in \cite{Deng&Dong:ICC22}.}
\thanks{Yong Deng is with the Department of Software Engineering, Lakehead University, Thunder Bay, ON P7B 5E1, Canada. He was with the Edward S. Rogers Sr. Department of Electrical and Computer Engineering,
University of Toronto, Toronto, ON M5S 3G4, Canada (email:
yong.deng@lakeheadu.ca).}
\thanks{Min Dong is with the Department of Electrical, Computer and Software Engineering, Ontario Tech University, Oshawa, ON L1G 0C5, Canada (e-mail: min.dong@ontariotechu.ca).}}


\allowdisplaybreaks
\IEEEtitleabstractindextext{\begin{abstract}
This paper studies MapReduce-based heterogeneous coded distributed computing (CDC) where, besides different computing capabilities at workers, input files  to be accessed by computing jobs have  nonuniform popularity. We propose a file placement strategy  that can handle an arbitrary number of input files. Furthermore,  we design  a nested coded shuffling strategy that can efficiently manage the nonuniformity of file popularity to maximize  the coded multicasting opportunity. We then formulate the joint optimization of   the proposed file placement  and nested shuffling design variables to optimize the proposed CDC scheme. To reduce the high computational complexity in solving the resulting mixed-integer linear programming (MILP) problem, we propose a simple two-file-group-based  file placement approach to obtain an approximate solution. 
Numerical results show that the optimized CDC scheme outperforms other alternatives. Also, the proposed two-file-group-based  approach achieves nearly the same performance as   the conventional branch-and-cut method in solving the MILP problem but with substantially lower  computational complexity that is scalable over the number of files and workers.  For computing jobs with aggregate target functions that commonly appear  in machine learning applications, we  propose a heterogeneous compressed CDC (C-CDC) scheme  to further improve the shuffling efficiency. The C-CDC scheme uses a local data aggregation technique to compress the data to be shuffled  for the shuffling load reduction. We again optimize the proposed C-CDC scheme and explore the two-file-group-based low-complexity approach for an approximate solution. Numerical results show the proposed C-CDC scheme provides a
considerable shuffling load reduction over the CDC scheme, and also, the two-file-group-based file placement approach maintains good performance.

\end{abstract}
 }

\maketitle

\allowdisplaybreaks
\section{Introduction}
\label{sec:intro}
Large-scale distributed computing has emerged as a rapidly growing computing paradigm that caters to the growing needs of large-scale machine learning and data analytics for data-intensive applications~\cite{Shanahan:SIGKDD15}. Popular distributed computing models, including MapReduce~\cite{Dean2008:ComACM,Sakr:ACMCOM13} and Spark~\cite{Shanahan:SIGKDD15}, employ the Map-Shuffle-Reduce framework to execute a job, where a job is split into several target functions, each executed via Map, Shuffle, and Reduce phases. In the Map phase, each worker uses its local data (files) to compute a set of map functions to generate intermediate values (IVs). In the Shuffle phase, workers exchange their locally generated IVs to acquire all the IVs they need to compute the target functions. In the Reduce phase, each worker computes its designated target functions by utilizing the local IVs and the IVs collected from the Shuffle phase. 

The major performance bottleneck of distributed computing comes from the Shuffle phase: A large number of IVs need to be shuffled among the workers, causing a high communication load. For example, it has been shown that when computing  a Tarasort task on Amazon EC2 platform, more than $90\%$ of the execution time is committed to data shuffling~\cite{Li&Maddah-Ali:CommMag17}. To alleviate the communication bottleneck in the Shuffle phase, the seminal
work~\cite{Li&Maddah-Ali:TIT2018} proposed a coded distributed computing (CDC) scheme, which reduces the data shuffling time substantially by utilizing  coded multicasting for exchanging IVs in the Shuffle phase.
For a homogeneous system, it is further shown that the proposed CDC scheme characterizes the exact computation-communication tradeoff for distributed computing~\cite{Li&Maddah-Ali:TIT2018}. Since then,  many efforts have been devoted to extending the CDC scheme to wireless systems~\cite{Li&Yu:TNET17,Li:TIT19,Shakya:2022On} and various computing scenarios, such as
 jobs
with linear functions~\cite{Elkordy&Li2021}, cascaded CDCs where a reduce
function is assigned to multiple workers~\cite{Woolsey:2021A2,Jiang:2023Cascaded,cheng2023asymptotically}, and CDC with stragglers~\cite{Yan:2020A}.
The works mentioned above commonly assume a homogeneous system with uniform computing capabilities among workers. However, the workers in practical systems may have diverse computing capabilities. This has motivated a great interest in recent works in designing CDC when the workers have different mapping loads and reducing loads~\cite{Kiamari&Wang:GCOM17,Xu&Tao:GCOM19,Xu&Shao:TCOM21,Woolsey:2021A,Song:2022Joint}, which are referred to as the heterogeneous CDCs.  

In all the above works, a common assumption for designing the CDC schemes is that all (input) files are needed by each job. In particular, the existing designs mainly focus on the execution of a single job, which requires access to all the files. In practice, however, skewed data file popularity is a typical characteristic  in the distributed computing systems~\cite{Anan2011:EuroSys}. Empirical analysis on industrial
MapReduce clusters (such as Facebook, Cloudera and Microsoft Bing's data center)
has shown that the file access pattern demonstrates a Zipf distribution: A small number of files account for most of  the file accesses by different jobs~\cite{Anan2011:EuroSys}.  For designing a conventional uncoded distributed computing scheme, it is shown that understanding the file popularity is essential to achieve a lower shuffling load~\cite{Anan2011:EuroSys,Chen&Yao:TPDS2015,Chen&Liu:TEM2021}.  Similarly, the existing CDC strategies are not suitable in this scenario. In particular, with different file popularity, the file subsets placed at the workers may now have variable sizes, while the existing schemes for exchanging IVs in the Shuffle phase are not designed to handle such a situation and thus result in inefficiency for shuffling.
 To our best knowledge, no existing work has considered nonuniform file popularity in designing CDC. In addition, due to the design requirements, the existing CDC schemes are only applicable when the number of files takes certain specific values, which typically need to be sufficiently large~\cite{Yan:2020A,Woolsey:2021A2,Jiang:2023Cascaded,cheng2023asymptotically,Elkordy&Li2021,Li&Maddah-Ali:TIT2018,Li:TIT19,Li&Yu:TNET17,Shakya:2022On,Kiamari&Wang:GCOM17,Xu&Tao:GCOM19,Woolsey:2021A,Xu&Shao:TCOM21},  limiting the use of these CDC schemes in practice.

Motivated by the above discussion, in this work, we  consider the heterogeneous CDC, where besides the commonly considered non-equal mapping loads and reducing loads among workers, the set of files  required for job access  are of arbitrary size and have nonuniform popularity.  We design the heterogeneous CDC scheme under this more general system model. Our contribution is summarized below. 
\begin{itemize}
\item 
In the Map phase, we propose a file placement strategy that is capable of accommodating
any number of files  with nonuniform popularities. Due to nonuniform file popularity,  the number of IVs to be shuffled at different workers are different, creating coding difficulty for shuffling. To overcome this challenge, in the Shuffle phase, we propose a nested coded shuffling strategy that can handle variable sizes of IVs efficiently and maximize the coded multicasting opportunity to reduce the shuffling load. 
\item
We further optimize the above proposed heterogeneous CDC scheme.  Specifically, we formulate a joint optimization of  
the  file placement and  the nested coded shuffling design variables in the proposed CDC scheme to minimize the expected
shuffling load.
The optimization problem is a mixed-integer linear programming (MILP)
problem, which is NP-hard. Using the existing branch-and-cut method~\cite{tahernejad2020branch} to obtain an approximate solution incurs high computational complexity. Thus,  we develop a simple low-complexity two-file-group-based approach to find an approximate solution. In particular, we propose a two-file-group-based
file placement strategy, where files are partitioned into popular and unpopular file groups, and different placement strategies are designed for each group for placing files at the workers. Under this file placement strategy, we reduce the original optimization problem to a coded shuffling
optimization problem, which is shown to be a linear
programming (LP) problem. The optimal file group partition can be obtained by a simple search. 
\item
We further improve the efficiency of our proposed heterogeneous CDC scheme for the class of computing
jobs with aggregate target functions, which are common in machine learning problems. We propose a heterogeneous compressed CDC
(C-CDC) by adding an extra IV aggregation step in the Shuffle phase of the previously proposed CDC scheme.  In particular, the IVs are aggregated locally at each worker before shuffling, such that only the compressed contents are shuffled to further reduce the shuffling load. We again optimize the file placement and the shuffling design variables in the proposed C-CDC scheme to minimize the shuffling load. We show that the previously proposed low-complexity two-file-group-based approach can also be applied to obtain an approximate solution for the optimized C-CDC scheme. 
\item
Numerical studies show that our proposed optimized CDC scheme can flexibly handle an arbitrary number of files and achieve a lower shuffling load than the existing schemes.   For the computing jobs with aggregate target functions, our results also show  that the proposed C-CDC scheme provides a considerable shuffling load reduction over the CDC scheme.
Furthermore, our numerical results demonstrate that the  proposed two-file-group-based approach performs very close to the branch-and-cut method for optimizing the CDC and C-CDC schemes but with orders of magnitude of lower computational complexity. 
\end{itemize}


\emph{Organization:} The rest of the paper is organized as follows. We first discuss the related works in Section~\ref{sec:related}. In Section~\ref{sec:sys}, we present the system model. We propose the heterogeneous
CDC scheme for nonuniform file popularity in Section~\ref{sec:problem}. Then, we  optimize the proposed CDC scheme in Section~\ref{sec:placement}. In Section~\ref{sec:C-CDC}, we further propose the heterogeneous
C-CDC for computing jobs with aggregate target functions to improve shuffling efficiency and provide our methods to optimize it.
The numerical studies are conducted in Section~\ref{sec:simu}, followed by the conclusion in Section~\ref{sec:conclusion}.

\emph{Notations:} The cardinality of set $\Sc$ is denoted by $|\Sc|$. Notation
 $\Ac\backslash\Sc$ denotes  subtracting the elements in set  $\Sc$ from
set $\Ac$. The bitwise "XOR" operation between two IVs is denoted by $\oplus$. 

\section{Related Works}\label{sec:related}
The CDC paradigm was first introduced in~\cite{Li&Maddah-Ali:TIT2018} to address the high shuffling load faced by the MapReduce-based distributed computing system. The proposed scheme in~\cite{Li&Maddah-Ali:TIT2018} applies coded multicasting, a technique originally proposed to alleviate the high traffic load in cache-aided
communication~\cite{Maddah-Ali&Niesen:TIT2014,ji2016fundamental,Zhang&Coded:TIT18,Ji&Order:TIT17,Deng&Dong:TIT22,Deng&DongMCCS:TIT22}, to the Shuffle phase of distributed computing.
It is shown that the shuffling load under the proposed CDC scheme is significantly lower than that of the traditional uncoded distributed computing in a homogeneous system, where all the workers have the same mapping and reducing loads, and all files need to be accessed by each job. 

\subsection{CDC for Homogeneous Systems}
The CDC has been further investigated for improvement. 
In~\cite{Li&Yu:TNET17,Li:TIT19,Shakya:2022On}, the design of CDC for wireless systems has been studied, where the workers communicate with each other via a central access point~\cite{Li&Yu:TNET17} or through an interference network~\cite{Li:TIT19,Shakya:2022On}. In~\cite{Elkordy&Li2021}, a compressed CDC
was proposed  to speed up the computation of jobs with
linear aggregation functions, which are commonly seen in machine learning applications. In~\cite{Woolsey:2021A2,Jiang:2023Cascaded,cheng2023asymptotically},
a cascaded CDC framework is considered where each target function
may be computed by multiple workers to provide fault tolerance for the system.
The CDC with stragglers is considered in~\cite{Yan:2020A}, where the workers may become unavailable randomly, and the storage-communication tradeoff is characterized for this scenario. The above works~\cite{Li&Maddah-Ali:TIT2018,Li&Yu:TNET17,Shakya:2022On,Li:TIT19,Elkordy&Li2021,Yan:2020A,Woolsey:2021A2,Jiang:2023Cascaded,cheng2023asymptotically}
all focus on a homogeneous scenario similar to~\cite{Li&Maddah-Ali:TIT2018}, where the workers have the same mapping
and reducing load, while all files are needed by each job.

\subsection{Heterogeneous CDC}

 Since workers typically have various computing capabilities, in practice,  heterogeneous CDC that considers nonuniform mapping load and/or reducing load has been investigated by many recent works~\cite{Kiamari&Wang:GCOM17,Xu&Tao:GCOM19,Woolsey:2021A,Xu&Shao:TCOM21,Song:2022Joint}. In~\cite{Kiamari&Wang:GCOM17}, a three-worker system was considered  where workers have uniform reducing load but nonuniform mapping loads. The authors proposed an optimal file placement strategy that achieves the minimum shuffling load. The works in~\cite{Xu&Tao:GCOM19,Woolsey:2021A,Xu&Shao:TCOM21,Song:2022Joint} focused on the design of  the CDC when both mapping and reducing loads are nonuniform. In~\cite{Xu&Tao:GCOM19}, the authors proposed a file-partitioning-based file placement strategy and optimized it jointly with the target function assignment to minimize the shuffling load. In~\cite{Xu&Shao:TCOM21}, a lower bound and an upper bound on the shuffling load have been developed for the heterogeneous CDC. In particular, the achievable upper bound is   obtained by jointly optimizing the file placement and coded shuffling strategy. In~\cite{Woolsey:2021A},
a heterogeneous CDC scheme has been constructed by using a heuristic combinatorial method for the design of both the coded shuffling   and the target function assignment. By comparing with the lower bound on shuffling load, the proposed scheme is
shown to be within a constant factor away from the optimal~\cite{Woolsey:2021A}. The work in~\cite{Song:2022Joint} considered the jointly optimal design of coded shuffling and target function assignment for the heterogeneous CDC. Since obtaining the optimal solution is computationally prohibitive,  a greedy algorithm is developed to find a feasible solution. Other than nonuniform mapping and reducing loads, \cite{Shakya:2022On} considered CDC with  heterogeneous communication capacities in wireless systems, and \cite{Dong&Tang:20ICPADS} studied the CDC design with heterogeneous IV sizes. 
In all the above works, although different heterogeneities have been studied, they are all under the common assumption that files are equally popular and each job needs to access all the files.

However, the nonuniformity of file popularity is a key characteristic in practice for distributed computing. Traditionally, it has been explored to speed up the uncoded distributed computing systems~\cite{Anan2011:EuroSys,Chen&Yao:TPDS2015,Chen&Liu:TEM2021}. In~\cite{Anan2011:EuroSys}, a distributed computing framework named Scarlett was proposed to utilize the knowledge of file popularity distribution in the file placement design for uncoded MapReduce, which was shown to speed up job execution by $20\%$ in a Hadoop cluster. In~\cite{Chen&Yao:TPDS2015}, the authors studied the target  function assignment under nonuniform file popularity distribution and proposed a lightweight implementation of a balanced range assignment strategy, which was shown to improve the job execution time in MapReduce by up to a factor of $4$. In~\cite{Chen&Liu:TEM2021}, the file placement problem under nonuniform file popularity was studied under a tree-topology network model. These results demonstrate that considering the file popularity for distributed computing is essential to improve  performance. The design of CDC under nonuniform file
popularity is much more challenging due to the complication of using coded multicasting for data
shuffling, and none of the existing works has considered this scenario. In this work, we provide our design of heterogeneous CDC that effectively takes into account the nonuniformity of the file popularity.

A part of this work has appeared in~\cite{Deng&Dong:ICC22}, which only presents our preliminary studies on the heterogeneous CDC design.
In the current paper, we further  propose a new heterogeneous C-CDC scheme for the class of computing jobs with aggregate target functions, which greatly improves the shuffling efficiency over the original CDC scheme. Moreover, we consider different approaches to optimize the proposed C-CDC scheme. Our extension of the prior work also provides an improved structure of the proposed shuffling strategy,  new derivation, complete proof, computational complexity analysis of the proposed algorithms, and more comprehensive numerical results.

\section{System Model}\label{sec:sys}

We consider a distributed computing framework with a MapReduce structure, where the network processes jobs from a set of $N$ files using $K$ distributed workers. Let $\Kc\triangleq\{1,\ldots,K\}$ and $\Nc\triangleq\{1,\ldots,N\}$ denote the worker and file indexes, respectively.
Each file $n\in\Nc$ is of size $F$ bits. We assume that these files may have different popularities to be used by the jobs,  measured by the frequency of a file being  accessed by the jobs. Such nonuniform popularity of files is common in practical systems~\cite{Anan2011:EuroSys,Chen&Yao:TPDS2015,Chen&Liu:TEM2021}. Let $\pbf=[p_1,\ldots,p_N]$ denote the popularity distribution vector of the $N$ files, where $p_n$ is the probability of file $n$ being accessed by a job, and $\sum_{n=1}^{N}p_n=1$. Without loss of generality, we assume the files are indexed according to the decreasing order of their popularities, s.t. $p_1\ge\ldots\ge p_N$. The $K$ workers are expected to accomplish each job cooperatively. The processing of a job requires accessing to a subset of files $\Dc\subseteq\Nc$, where $\Dc \neq \emptyset$. We refer $\Dc$ as the files for the job, and $D=|\Dc|$ is the number of files in $\Dc$. 

By the MapReduce framework, each job is split into $Q$ target functions, denoted by $\phi_1(\Dc),\ldots,\phi_Q(\Dc)$. Each
target function $\phi_q(\Dc)$ maps all the files in $\Dc$ to an output stream $\Phi_q$ of $B$ bits: $\Phi_q=\phi_q(\Dc)$. These $Q$ target functions are distributed to different workers to compute.
Let $\Qc\triangleq\{1, \ldots, Q\}$ denote the index set of the target functions.

\begin{figure}
\psfrag{IVs}{\hspace{-1.2em}IVs}
\psfrag{V_{1,1}}{\tiny $V_{1,1}$}
\psfrag{V_{1,n}}{\tiny $V_{1,n}$}
\psfrag{V_{1,D}}{\tiny $V_{\!1,\!D}$}
\psfrag{V_{q,1}}{\tiny $V_{q,1}$}
\psfrag{V_{q,n}}{\tiny $V_{q,n}$}
\psfrag{V_{q,D}}{\tiny $V_{\!q,\!D}$}
\psfrag{V_{Q,1}}{\tiny $V_{\!Q,1}$}
\psfrag{V_{Q,n}}{\tiny $\!V_{\!Q,n}$}
\psfrag{V_{Q,D}}{\tiny $\!V_{\!Q,\!D}$}

\psfrag{Map phase}{\small \hspace{-.0em}Map}
\psfrag{g_{1,n}}{\small$g_{1,n}$}
\psfrag{g_{q,n}}{\small$g_{q,n}$}
\psfrag{g_{Q,n}}{\small$\!g_{Q\!,n}$}

\psfrag{g_{1,1}}{\small$g_{1,1}$}
\psfrag{g_{q,1}}{\small$g_{q,1}$}
\psfrag{g_{Q,1}}{\small$g_{Q\!,1}$}

\psfrag{g_{1,D}}{\small$g_{1\!,\!D}$}
\psfrag{g_{q,D}}{\small$g_{q\!,\!D}$}
\psfrag{g_{Q,D}}{\small$g_{Q\!,\!D}$}

\psfrag{Reduce phase}{\small \hspace{-.0em}Reduce}
\psfrag{h_1}{\small $\!\!\!\!h_1\!(\!\{\!V_{1\!,\!1}, \!...,\! V_{1\!,\!D}\!\}\!)$}
\psfrag{h_q}{\small $h_q\!(\!\{\!V_{q\!,\!1},\! ...,\! V_{q\!,\!D}\!\}\!)$}
\psfrag{h_Q}{\small $h_Q\!(\!\{V_{Q\!,\!1},\! ...,\! V_{Q\!,\!D}\!\}\!)$}

\psfrag{phi_1}{\small$\phi_1(\Dc)$}
\psfrag{phi_q}{\small$\phi_q(\Dc)$}
\psfrag{phi_Q}{\small$\phi_Q(\Dc)$}

\psfrag{=}{$=$}

\psfrag{1}{$1$}
\psfrag{n}{$n$}
\psfrag{D}{$D$}
\psfrag{Target functions}{Target functions}

\psfrag{Job}{\hspace{-2.4em}A job requiring files $\Dc$}
\psfrag{Input file set D}{\hspace{-4.2em}file set $\Dc=\{1,\ldots,n,\ldots,D\}$}
\includegraphics[width=1\linewidth]{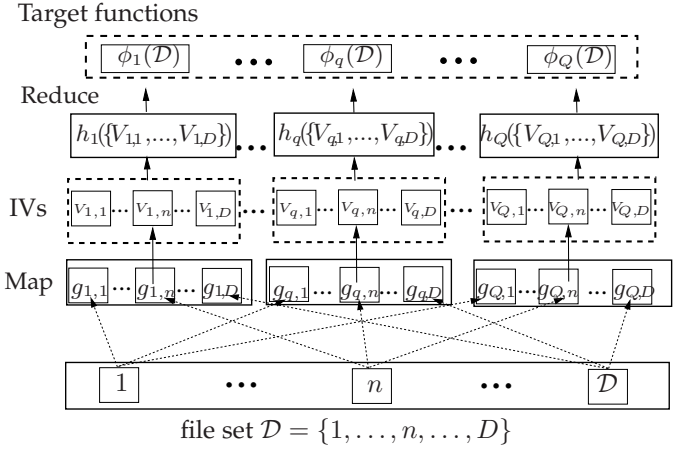}
\caption{Processing of of a job in the MapReduce framework. The set of files required by the job is $\Dc$.}\label{fig:mapreduce}
\end{figure}

Each target function $\phi_q(\Dc)$, $q\in\Qc$, is computed via a composition of Map and Reduce functions. Specifically, each target function has $D$ Map functions, one for each file $n\in\Dc$, denoted by $g_{q,n}$. There are a total of $QD$ target functions for each job. For target function $\phi_q(\Dc), q\in \Qc$, each Map function $g_{q,n}$ outputs an Intermediate Value (IV) of $T$ bits, denoted by $V_{q,n}$, for $n\in\Dc$. Each target function $\phi_q(\Dc), q\in\Qc$, is then computed by a Reduce function $h_q$  using the corresponding IVs $\{V_{q,n}, n\in\Dc\}$ generated from the $D$ Map functions, \ie
\begin{align}\label{equ:reduce_func}
\phi_q(\Dc)=h_q(\{V_{q,n}, n\in\Dc\}),\quad  q\in\Qc.
\end{align}
The above procedure of processing a job through the MapReduce structure is illustrated in Fig.~\ref{fig:mapreduce}.

To process each job among $K$ workers, each worker $k$ computes a subset of the $QD$ Map functions based on its stored files to generate its local IVs.  Let $M_k\in\mathbb{N}^+$ denote the  number of files that can be stored by worker $k$, which is referred to as the mapping load. Following the common practice~\cite{Li&Maddah-Ali:TIT2018,Kiamari&Wang:GCOM17,Xu&Tao:GCOM19,Xu&Shao:TCOM21,Dong&Tang:20ICPADS}, we assume $\sum_{k=1}^{K}M_k\ge N$, which guarantees that all the files can be stored among the $K$ workers. Let $\Mc_k\subseteq\Nc$ denote the set of files stored at worker $k$. Based on these available files, each worker $k$ computes a set of Map functions for each target function $\phi_q(\Dc)$, using the stored files in $\Mc_k\cap\Dc$, given by $\{g_{q,n}, n\in\Mc_k\cap\Dc, q\in\Qc\}$.

For processing each job, the $Q$ target functions are distributed to $K$ workers to compute. In this work, we consider a general target function assignment scenario, where each target function is assigned to only one worker and each worker may be assigned a different number of target functions. Let $\Wc_k$ denote the set of target functions assigned to worker $k$, where $W_k\triangleq|\Wc_k|/Q$, as a fraction of the total $Q$ target functions, is the reducing load of worker $k$. Since each target function is assigned to only one worker, we have $\sum_{k=1}^{K}W_k=Q$.

In summary, the distributed computing network computes any job in three phases: \emph{Map, Shuffle,} and \emph{Reduce}.
\begin{itemize}
\item
In the Map phase, each worker $k\in\Kc$ computes a st of Map functions for all the $Q$ target functions, according to its stored files $\Mc_k\cap\Dc$, to generate its local IVs, given by $\{V_{q,n}: n\in\Mc_k\cap\Dc, q\in\Qc\}$.
\item
In the Shuffle phase, each worker $k$ generates message $X_k$ using its local IVs and multicasts it to all the other workers. Each worker $k$ receives the multicasted messages from the other workers, $\{X_i,i\in\Kc\backslash\{k\}\}$, to obtain the required IVs for its assigned target functions in $\Wc_{k}$ that are not already computed locally,  \ie$\{V_{q,n}: q\in\Wc_k,n\notin\Mc_k, n\in\Dc\}$.
\item
In the Reduce phase, each worker $k$ computes its assigned target functions in $\Wc_k$ through the Reduce functions, using its local IVs  $\{\phi_q(\Dc) = h_q(\{V_{q,n}, n\in\Dc\}), q\in\Wc_k\}$ and other required IVs obtained in the Shuffle phase.
\end{itemize}

{The notations used in the paper are summarized in Table~\ref{tab:notation}.}

\section{Heterogeneous Coded Distributed Computing}\label{sec:problem}

The original CDC scheme~\cite{Li&Maddah-Ali:TIT2018} is constructed for a single job, under the assumption of a sufficiently large number of files ($N$) and assuming uniform mapping load and reducing load among workers. The main idea there is to form coded messages to explore multicasting opportunities in the Shuffle phase.
Specifically, each worker is assigned the same number of target functions.  For placement of files at workers in the Map phase, the files are partitioned into file subsets. A unique file subset  (which can be empty) is assigned to a unique worker subset $\Sc\subseteq \Kc$, and the size of file subsets is the same for the worker subsets of the same size.   Each worker $k\in\Sc$ needs  IVs computed by the other workers in $\Sc\backslash\{k\}$, and and the number of such IVs is the same for any worker $k\in\Sc$. In the Shuffle phase, each worker $k\in\Sc$ generates a coded message containing the IVs needed by all the rest workers in $\Sc$  and multicast the message to them. Note that this scheme requires $N$ to be large enough for its file placement, such that the files can be partitioned into required subsets, which limits its applicability in the general system settings.

In this work, we consider a heterogeneous CDC scenario where $N$ files has nonuniform popularity of being accessed by jobs. Furthermore, in contrast to the above original CDC scheme, we propose a file placement strategy for any number $N$ of files. With file popularity being different, the file subsets placed at the worker subsets of the same size may now be of different sizes. Thus,  the number of IVs needed by different workers in the same worker subset may be different, resulting in variable lengths of messages at each worker that need to be coded together for multicasting. Consequently, some  IVs  cannot be encoded into the same coded message. One straightforward solution to handle this issue is to unicast these remaining IVs. However, it leads to a performance loss due to not taking advantage of coded multicasting opportunities~\cite{Kiamari&Wang:GCOM17,Xu&Tao:GCOM19}. To overcome this inefficiency, we   propose a nested coded shuffling strategy to encode all IVs. Intuitively, this strategy  exploits extra coded multicasting opportunities for those remaining IVs, leading to reduce the shuffling load as compared with the unicast solution.  
\renewcommand{\arraystretch}{1.3}
\begin{table}
\renewcommand{\arraystretch}{1}
   \centering
   \caption{{Summary of Key Notations}}
   \label{tab:notation}
   \vspace{-1em}
   \resizebox{\columnwidth}{!}{
   \begin{tabular}{ll}
   \toprule\toprule
   \textbf{Symbol} & \hspace{.0em}\textbf{Definition}\\
   \midrule
   $\Qc$ & Set of all the target functions\\ \midrule
   $Q$ & Size of $\Qc$\\ \midrule
   $\Dc$ & Set of the files being requested by the jobs\\ \midrule
   $\Sc_{+i}$ & Subset $\Sc\cup\{i\}$\\ \midrule
   $\Sc_{-i}$ & Subset $\Sc\backslash\{i\}$\\ \midrule
   $\Mc_k$ & Set of files stored by worker $k$\\ \midrule
   $M_k$ & Size of $\Mc_k$\\ \midrule
   $\Wc_k$ & Set of target function assigned to worker $k$\\ \midrule
   $W_k$ & Size of $\Wc_k$\\ \midrule
   $t_{n,\Sc}$ & The indicator of whether file $n$ is placed in subset $\Sc$\\ \midrule
   $\Ac_{\Sc}^\Dc$ & Set of files  $\Dc$ placed exclusively in subset $\Sc$\\ \midrule
   $a_{\Sc}^\Dc$ & Size of $\Ac_{\Sc}^\Dc$\\ \midrule
   $\Vc^\Dc_{k,\Sc_{-\!k}}$ & IVs needed by worker $k$,  computed by workers in $\Sc_{-\!k}$\\ \midrule
 $\Rc_{k,\Sc_{+\!i}}^{\Sc}\!(\Dc)$ & Residual IVs needed by worker $k$, but not coded in $\Sc_{+\!i}$\\ \midrule
$r_{k,\Sc_{+\!i}}^{\Sc}\!(\Dc)$ & Normalized size of $\Rc_{k,\Sc_{+\!i}}^{\Sc}\!(\Dc)$\\ \midrule
\multirow{2}{1.cm}{$\Upsilon_{k,\Sc}^{\Dc}$} & All residual IVs needed by worker $k$ that are\\ &from $\Sc_{+i}$'s, for $i\in\Kc\backslash\Sc$\\ \midrule
$\Ic_{k,\Sc}^\Dc$ & Set of  IVs to be sent to worker $k$ within $\Sc$\\ \midrule
$\Ic^{\Dc}_{k,\Sc|j}$ & Portion of $\Ic_{k,\Sc}^\Dc$ that is locally computed by worker $j\in\Sc$\\ \midrule
$L_{j,\Sc}^\Dc$ & Normalized size of $\Ic^{\Dc}_{k,\Sc|j}$, $\forall k\in\Sc\backslash\{j\}$\\ \midrule
$C_{\Sc|j}^{\Dc}$ & Coded message using $\Ic^{\Dc}_{k,\Sc|j}$'s, for  $k\in\Sc\backslash\{j\}$\\ \midrule 
$\widetilde\Vc^\Dc_{k,\Sc_{-\!k}}$ & Aggregated IVs needed by worker $k$,  computed by $\Sc_{-\!k}$\\ \midrule
 \multirow{2}{1.cm}{$\widetilde\Rc_{k,\Sc_{+\!i}}^{\Sc}\!(\Dc)$} & Residual aggregated IVs needed by worker $k$, \\ 
&but not coded in $\Sc_{+\!i}$\\ \midrule
$\widetilde r_{k,\Sc_{+\!i}}^{\Sc}\!(\Dc)$ & Normalized size of $\widetilde\Rc_{k,\Sc_{+\!i}}^{\Sc}\!(\Dc)$\\ \midrule
$\widetilde\Ic_{k,\Sc}^\Dc$ & Set of aggregated  residual IVs to be sent to worker $k$ within $\Sc$\\ \midrule
$\widetilde\Ic^{\Dc}_{k,\Sc|j}$ & Portion of $\widetilde\Ic_{k,\Sc}^\Dc$ that is locally computed by worker $j\in\Sc$\\ \midrule
$\widetilde L_{j,\Sc}^\Dc$ & Normalized size of $\widetilde\Ic^{\Dc}_{k,\Sc|j}$, $\forall k\in\Sc\backslash\{j\}$\\ \midrule
$\widetilde C_{\Sc|j}^{\Dc}$ & Coded message using $\widetilde\Ic^{\Dc}_{k,\Sc|j}$'s, for  $k\in\Sc\backslash\{j\}$\\ \midrule
\bottomrule
   \end{tabular}}
\end{table}

\subsection{The Mapping Strategy}\label{sec:mapping}

In the Map phase, the files are placed at the workers, and the Map functions are computed to generate local IVs.  {For $K\ $workers, there are $2^{K}-1$ non-empty worker subsets in $\Kc$: $\{\Sc:\Sc\subseteq\Kc,\Sc\ne\emptyset\}$. When a file is placed at a set of workers (one of  $2^K-1$ subsets), we can use  $\Sc$ to index its placement at these workers. In other words, each of the $N$ files is placed in only one \emph{unique} worker subset $\Sc$ among the $2^{K}-1$ subsets. Let $t_{n,\Sc}$ be a file placement indicator to indicate whether file $n$ is exclusively placed in the worker subset $\Sc$}~\footnote{{Note that the file placement strategy associates each file with only one unique worker subset in $\{\Sc:\Sc\subseteq\Kc,\Sc\ne\emptyset\}$, which contains \emph{all} workers that the file is stored at. Although for a file placed at $\Sc$, it is also located in any subset $\Sc^{\prime}\subset\Sc$,  the subset $\Sc^{\prime}$ does not contain all workers that the file is stored at and is incomplete. Based on our file placement strategy description, it is only correct to say that the file is placed in $\Sc$.}}:
\begin{align}\label{eq:con1}
t_{n,\Sc}\in\{0,1\}, \quad n\in\Nc, \Sc\subseteq\Kc, \Sc \neq \emptyset.
\end{align}
 Then, we have the file placement constraint
for each file as \begin{align}\label{eq:con2}
 \sum_{\Sc\subseteq\Kc, \Sc \neq \emptyset}t_{n,\Sc}=1, \quad\  n\in\Nc.
 \end{align}
Furthermore, the files placed at worker $k\in\Kc$ cannot exceed the  worker's mapping load $M_k$, given by the following mapping load constraint
\begin{align}\label{eq:con3}
  \sum_{n=1}^{N}\ \sum_{\Sc\subseteq\Kc, \Sc \neq \emptyset,k\in\Sc}t_{n,\Sc}\le M_k, \quad\ k\in\Kc.
\end{align}
Consider a job requiring the set of files $\Dc$. Let $\Ac_{\Sc}^\Dc\subseteq\Dc$ denote the set of files placed exclusively at  worker subset $\Sc$, and let $a_{\Sc}^\Dc=|\Ac_{\Sc}^\Dc|$. Then, we have
\begin{align}\label{equ:a_S_D}
a_{\Sc}^\Dc=\sum_{n\in\Dc}t_{n,\Sc}, \quad \Sc\subseteq\Kc, \Sc \neq \emptyset.
\end{align}

Let $\Tc\triangleq\{ t_{n,\Sc}: n\in \Nc, \Sc\subseteq\Kc, \Sc \neq \emptyset\}$ represent the file placement strategy. Based on $\Tc$, the workers compute the Map functions to generate local IVs. The IVs  generated exclusively by worker subset $\Sc$ are given by $\{V_{q,n}:q\in\Qc, n\in \Ac_{\Sc}^\Dc \}$.
We intend to optimize the file placement strategy to minimize the expected  shuffling load.

\begin{Remark}
We point out that the existing homogeneous and heterogeneous CDC schemes~\cite{Li&Maddah-Ali:TIT2018,Li:TIT19,Li&Yu:TNET17,Kiamari&Wang:GCOM17,Xu&Tao:GCOM19,Xu&Shao:TCOM21,Dong&Tang:20ICPADS}  all require the number of files being sufficiently large. Also, they assume each job requires all files and do not consider the heterogeneity of file popularity for different jobs.
In contrast, we propose a mapping strategy  for an arbitrary number of files with nonuniform popularity to be accessed by jobs.
\end{Remark}

\subsection{The Nested Coded Shuffling Strategy}\label{sec:shuffling}

In the Shuffle phase, each worker $k$ needs to obtain the IVs that are not computed locally, $\{V_{q,n}: q\in\Wc_k,n\notin\Mc_k, n\in\Dc\}$, in order to compute the assigned target functions.
As discussed earlier regarding a CDC scheme, in a worker subset, each worker encodes the IVs needed by other workers  in the subset into a coded message and multicasts it to them. However, in a heterogeneous scenario where  files have different popularities,
the number of the IVs needed by different workers in the same worker subset may be different. As a result, these IVs  cannot be directly coded into the same message, which complicates the design of coded shuffling strategy.

To tackle the above challenge and maximize the benefit of multicasting for reducing the shuffling load, we  propose a  nested coded shuffling strategy, which exploits coded multicasting opportunities for all IVs.
Our strategy adopts an approach similar to the one proposed   in \cite{Xu&Shao:TCOM21} in processing a job. However, the heterogeneous CDC scenario considered in \cite{Xu&Shao:TCOM21} is different from ours, where   it is assumed that all files are needed to process a job, while the workers have different mapping loads and reducing loads. In our work, we consider a more general heterogeneous scenario, where each job may only require a subset of the $N$ files, and each file has different popularity of being accessed by jobs. 

Based on the file placement, each unique worker subset $\Sc$ is assigned an exclusive file subset. Thus, worker $k$ needs IVs from worker subsets in  $\{\Sc: \Sc \subseteq \Kc, k\notin \Sc\}$.
{In other words, worker $k$ needs IVs from all the other workers in any subset $\Sc$ containing worker $k$. We  first introduce a shuffling strategy for  worker $k$ to obtain IVs in a worker subset $\Sc\subseteq\Kc, |\Sc|\ge2$. It consists of three steps:}



\begin{itemize}
\item Step 1: \emph{Identify residual IVs.} Identify the IVs and the residual IVs to be sent to each  worker in $\Sc$;
\item Step 2: \emph{Nested grouping of IVs.} {Determine the portion of IVs to be coded into the same message  to be multicasted to other workers in  $\Sc$; Identify remaining IVs that are not coded in $\Sc$ as residual IVs to be sent in $\Sc^{\prime}\subset\Sc$;}
\item Step 3: \emph{Coded message formation.} Generate coded messages to be multicasted to other workers.
\end{itemize}

\begin{figure*}
\begin{subfigure}
\centering
\psfrag{I_{i,S}}{\hspace{-1.8em}{\small$\Ic_{k,\Sc}^\Dc, k= 1,2,3$}}
 \psfrag{I_{1,S}}{{\small$\Ic_{1,\Sc}^\Dc$}}
 \psfrag{I_{2,S}}{{\small$\Ic_{2,\Sc}^\Dc$}}
 \psfrag{I_{3,S}}{{\small$\Ic_{3,\Sc}^\Dc$}}
 
 \psfrag{I^D_{2,S|1}}{{\small$\Ic^{\Dc}_{2,\Sc|1}$}}
 \psfrag{I^D_{3,S|1}}{{\small$\Ic^{\Dc}_{3,\Sc|1}$}}
  
  \psfrag{I^D_{1,S|2}}{{\small$\Ic^{\Dc}_{1,\Sc|2}$}}
 \psfrag{I^D_{3,S|2}}{{\small$\Ic^{\Dc}_{3,\Sc|2}$}}
  
  \psfrag{I^D_{1,S|3}}{{\small$\Ic^{\Dc}_{1,\Sc|3}$}}
 \psfrag{I^D_{2,S|3}}{{\small$\Ic^{\Dc}_{2,\Sc|3}$}}
  
 \psfrag{R_{1,S}^{S_{-2}}(D)}{\hspace{-.5em}\small $\underset{i\in\Sc\backslash\{1\}}{\bigcup} \Rc_{1,\Sc}^{\Sc_{\!-\!i}}\!(\Dc)$}
 \psfrag{R_{2,S}}{\hspace{-.3em}\small $\underset{i\in\Sc\backslash\{2\}}{\bigcup} \Rc_{2,\Sc}^{\Sc_{\!-\!i}}\!(\Dc)$}
 \psfrag{R_{3,S}}{\hspace{-.3em}\small $\underset{i\in\Sc\backslash\{3\}}{\bigcup} \Rc_{3,\Sc}^{\Sc_{\!-\!i}}\!(\Dc)$}
 \psfrag{C_{S|1}}{$C^\Dc_{\Sc|1}$}
 \psfrag{C_{S|2}}{$C^\Dc_{\Sc|2}$}
 \psfrag{C_{S|3}}{$C^\Dc_{\Sc|3}$}

 \psfrag{xor}{$\bigoplus$}
 \psfrag{I_{3,{1,4}}}{$\Ic_{3,\{1,4\}}$}
  
 \psfrag{IVs needed by workers in S = 1,2,3,4 0}{\hspace{.8em}\footnotesize IVs needed by each}
\psfrag{IVs needed by workers in S = 1,2,3,4}{\hspace{.8em}\small worker in $\Sc=\{1,2,3\}$}
\psfrag{IVs coded and multicasted by workers in S 0}{\hspace{-.5em}\small IVs that are coded and multicasted}
\psfrag{IVs coded and multicasted by workers in S}{\hspace{-.5em}\small by workers in $\Sc$ as in \eqref{equ:codedmsg}}
\psfrag{Residual IVs to be considered in subsets of S }{\hspace{.0em}\small Residual IVs in $\Sc$}

\psfrag{From 1 0}{\hspace{-.0em}\footnotesize Needed by worker 1, to be }
\psfrag{From 1}{\footnotesize considered in $\{1,2\}$ and $\{1, 3\}$.}

\psfrag{From 2 0}{\hspace{-.0em}\footnotesize Needed by worker 2, to be }
\psfrag{From 2}{\footnotesize considered in $\{1,2\}$ and $\{2, 3\}$.}

\psfrag{From 3 0}{\hspace{-.0em}\footnotesize Needed by worker 3, to be }
\psfrag{From 3}{\footnotesize considered in $\{1,3\}$ and $\{2, 3\}$.}

\psfrag{=}{\LARGE$\bm{\Rightarrow}$}
\psfrag{-}{\LARGE$\bm{-}$}

\psfrag{R1_12}{\small$\!\Rc_{1,\Sc}^{\!\{\!1,2\!\}}\!(\!\Dc\!)$}
\psfrag{R1_13}{\small$\!\Rc_{1,\Sc}^{\!\{\!1,3\!\}}\!(\!\Dc\!)$}

\psfrag{R2_12}{\small$\!\Rc_{2,\Sc}^{\!\{\!1,2\!\}}\!(\!\Dc\!)$}
\psfrag{R2_23}{\small$\!\Rc_{2,\Sc}^{\!\{\!2,3\!\}}\!(\!\Dc\!)$}

\psfrag{R3_13}{\small$\!\Rc_{3,\Sc}^{\!\{\!1,3\!\}}\!(\!\Dc\!)$}
\psfrag{R3_23}{\small$\!\Rc_{3,\Sc}^{\!\{\!2,3\!\}}\!(\!\Dc\!)$}
\includegraphics[width=0.9\textwidth]{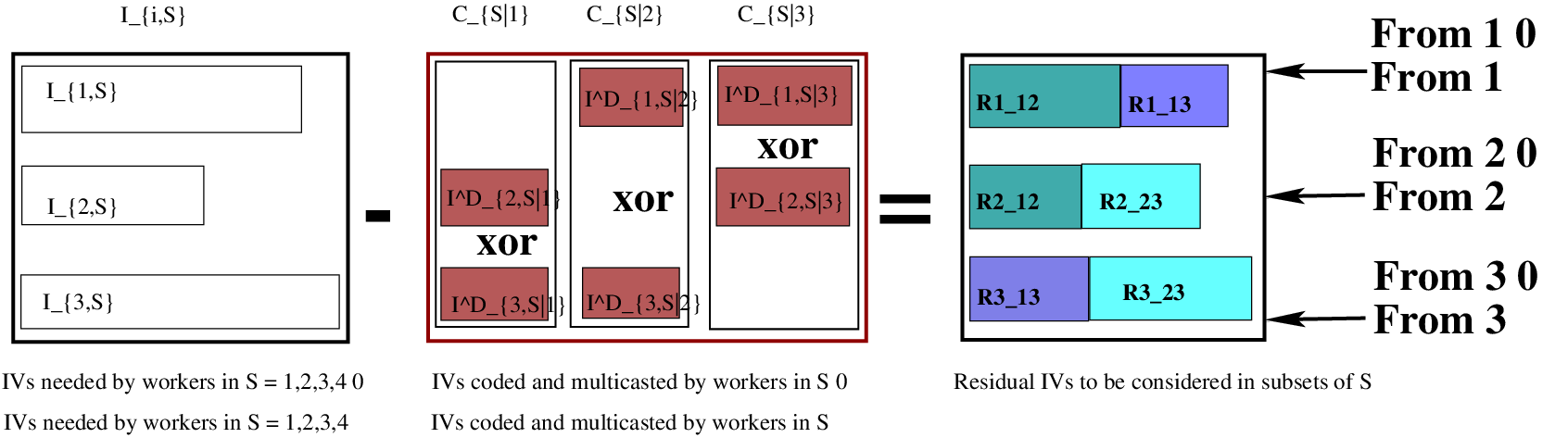}
\caption{{An example of Steps 2 and 3 in the nested coded shuffling strategy:  $\Sc=\Kc=\{1,2,3\}$. Left: IVs needed by the each worker in $\Sc$; Middle: IVs for the coded message by each worker; Right: Residual IVs to be considered in the subsets of $\Sc$.}}
\label{subfig1}
\end{subfigure}
\begin{subfigure}


\psfrag{R_{1,S1}}{\hspace{-.5em}\small $\Rc_{1,\Sc}^{\Sc^{\prime}}\!(\Dc)$}
\psfrag{U_R1}{\hspace{-5em}$\Upsilon_{1,\Sc^{\prime}}^{\Dc}=\bigcup_{i\in\Kc\backslash\Sc^{\prime}}\Rc_{1,\Sc^{\prime}_{+\!i}}^{\Sc^{\prime}}\!(\Dc)$}

\psfrag{Residual IVs from all the supersets of S1}{\hspace{-4em}{\small Residual IVs from $\Sc^{\prime}_{\!+i}, i\in\Kc\backslash\Sc^{\prime}$ that are requested by worker $1$}}

\psfrag{Local IVs 0}{\hspace{-1em}{\small Local IVs in}}
\psfrag{Local IVs}{\hspace{-1em}{\small $\Sc^{\prime}_{\!-\!1}=\{2,3\}$}}

\psfrag{=}{\large $\ \!\bm{=}$}
\psfrag{+}{\large $\bm{+}$}

\psfrag{IVs needed by workers in S1 = 1,2}{\hspace{.8em}\small IVs needed by workers in $\{1,2\}$}
\psfrag{IVs needed by workers in S2 = 1,3}{\hspace{.8em}\small IVs needed by workers in $\{1,3\}$}
\psfrag{IVs needed by workers in S3 = 2,3}{\hspace{.8em}\small IVs needed by workers in $\{2,3\}$}

\psfrag{Local IVs within S1}{\small Local IVs needed by worker $1$ and computed}
\psfrag{Local IVs within S12}{\small  by worker $2$ in $\{1,2\}$.}
\psfrag{Residual IVs from S, }{\small Residual IVs from $\Sc$, needed by worker $1$}
\psfrag{to be considered in S1}{\small and to be considered in $\{1,2\}$}

\psfrag{I_{1,S1}}{{\small$\Ic_{1,\{1,2\}}^\Dc$}}
\psfrag{V_{1,S1}}{{\small$\Vc_{1,\{2\}}^\Dc$}}
\psfrag{R_{1,S1}}{\hspace{-.5em}\small $\Rc_{1,\Sc}^{\!\{\!1,2\}}\!(\Dc)$}

\psfrag{I_{1,S2}}{{\small$\Ic_{2,\{1,2\}}^\Dc$}}
\psfrag{V_{1,S2}}{{\small$\Vc_{2,\{1\}}^\Dc$}}
\psfrag{R_{1,S2}}{\hspace{-.5em}\small $\Rc_{2,\Sc}^{\!\{\!1,2\}}\!(\Dc)$}

\psfrag{I_{1,S13}}{{\small$\Ic_{1,\{1,3\}}^\Dc$}}
\psfrag{V_{1,S13}}{{\small$\Vc_{1,\{3\}}^\Dc$}}
\psfrag{R_{1,S13}}{\hspace{-.5em}\small $\Rc_{1,\Sc}^{\!\{\!1,3\}}\!(\Dc)$}

\psfrag{I_{3,S13}}{{\small$\Ic_{3,\{1,3\}}^\Dc$}}
\psfrag{V_{3,S13}}{{\small$\Vc_{3,\{1\}}^\Dc$}}
\psfrag{R_{3,S13}}{\hspace{-.5em}\small $\Rc_{3,\Sc}^{\!\{\!1,3\}}\!(\Dc)$}

\psfrag{I_{2,S23}}{{\small$\Ic_{2,\{2,3\}}^\Dc$}}
\psfrag{V_{2,S23}}{{\small$\Vc_{2,\{3\}}^\Dc$}}
\psfrag{R_{2,S23}}{\hspace{-.5em}\small $\Rc_{2,\Sc}^{\!\{\!2,3\}}\!(\Dc)$}

\psfrag{I_{3,S23}}{{\small$\Ic_{3,\{2,3\}}^\Dc$}}
\psfrag{V_{3,S23}}{{\small$\Vc_{3,\{2\}}^\Dc$}}
\psfrag{R_{3,S23}}{\hspace{-.5em}\small $\Rc_{3,\Sc}^{\!\{\!2,3\}}\!(\Dc)$}

\includegraphics[width=0.95\textwidth]{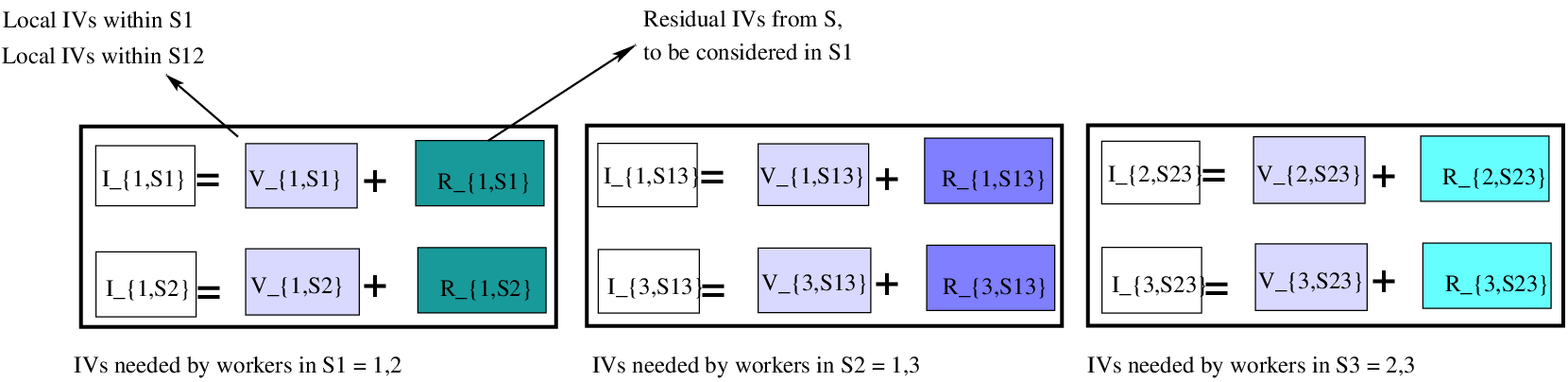}
\caption{{For the example in Fig.~\ref{subfig1}, the composition  of the IVs needed by each worker in all the subsets of $\Sc=\{1,2,3\}$ as in~\eqref{equ:I_CDC}.}}
\label{subfig2}
\end{subfigure}
\end{figure*}

{We will describe the three steps in detail below.
After that, we will describe the general nested coded shuffling strategy that applies this three-step shuffling strategy to all the worker subsets. To facilitate our description,  at the end of this section,  we use an example in Figs.~\ref{subfig1} and \ref{subfig2} to illustrate the procedure of our strategy in Steps  2-3. 
}

\subsubsection{Step 1: Identify Residual IVs}\label{sec:1step}  
 First, we identify the IVs that worker $k$ needs from other workers via shuffling. For any $\Sc \subseteq \Kc$, consider worker  $k\in\Sc $ and files placed in $\Sc_{-\!k}\triangleq \Sc\backslash\{k\}$. Worker $k$ needs to obtain  the IVs  computed by the  workers in $\Sc_{-\!k}$,  given   by 
\begin{align}\label{equ:IV_in_S}
\Vc^\Dc_{k,\Sc_{-\!k}}=\{V_{q,n}:q\in\Wc_k, n\in \Ac_{\Sc_{-\!k}}^\Dc \}
\end{align} 
Recall from \eqref{equ:a_S_D} that $a^\Dc_{\Sc_{-\!k}}=|\Ac^\Dc_{\Sc_{-\!k}}|$. Thus, the total size of  $\Vc^\Dc_{k,\Sc_{-\!k}}$ in bits is
\begin{align}\label{equ:size_V}
|\Vc^\Dc_{k,\Sc_{-\!k}}|=W_kTQa^\Dc_{\Sc_{-k}}=W_kTQ\sum_{n\in\Dc}\!t_{n,\Sc_{-\!k}}.
\end{align}
Note that  $|\Vc_{k,\Sc_{-\!k}}|$ may be different for each  $k \in \Sc$.

Next, for shuffling, we describe the IVs that each worker $j$ {needs to send to other workers. In particular, each worker $j\in\Sc$ needs to code the   IVs  needed by  other workers in $\Sc$  into a message. Note that the total size of the IVs needed by each worker may be different. Thus, worker $j$ takes a fraction of the IVs intended for each worker  $k\in\Sc\backslash\{j\}$, which has the same size for all these workers. These IVs will be coded into a message to be sent to $\Sc\backslash\{j\}$.  We will leave the details to  Sections~\ref{sec:2step}-\ref{sec:3step}. This procedure leaves some \emph{residual IVs} at worker $j$. 
{Let $\Sc_{+i}\triangleq \Sc\cup\{i\}$, for  $i\in\Kc\backslash\Sc$. Consider the above procedure  for worker subset $\Sc_{+i}$. Note that for $k\in\Sc$, we  have $k\in\Sc_{+i}$. Denote $\Rc_{k,\Sc_{+\!i}}^{\Sc}\!(\Dc)$ as the set of residual IVs that are
needed by worker $k$ in worker subset $\Sc_{+\!i}$ but are not coded into the message for $\Sc_{+\!i}$; Instead, they will be coded into messages within $\Sc$ (either $\Sc$ or some subsets of $\Sc$).} Define $r_{k,\Sc_{+\!i}}^{\Sc}\!(\Dc)\triangleq|\Rc_{k,\Sc_{+\!i}}^{\Sc}\!(\Dc)|/TQ$, which is the size of $\Rc_{k,\Sc_{+\!i}}^{\Sc}\!(\Dc)$ in bits normalized by $TQ$. \footnote{Following common practice~\cite{Li&Maddah-Ali:TIT2018,Li&Yu:TNET17}, we assume $T$ is large enough so that $TQ\sum_{i\in\Kc\backslash\Sc}\!r_{k,\Sc_{+\!i}}^{\Sc}\!(\Dc)\in\mathbb{Z}$.}
}

{Our proposed strategy is to  code the residual IVs  $\Rc_{k,\Sc_{+\!i}}^{\Sc}\!(\Dc)$ that remain from $\Sc_{+i}$ into  the messages for  $\Sc$ or the subsets of $\Sc$. In other words, those residual IVs  needed by worker $k$ from  all those supersets of $\Sc$ of size $|\Sc|+1$ will be encoded and sent in $\Sc$ or subsets of $\Sc$.
To prepare for this, we form the set of all residual IVs needed by worker $k$ that are collected from all the supersets of $\Sc$ of size $|\Sc|+1$, \ie $\Sc_{+\!i}$'s, for  $i\in\Kc\backslash\Sc$:} 
\begin{align}\label{equ:compo}
\Upsilon_{k,\Sc}^{\Dc}=\bigcup_{i\in\Kc\backslash\Sc}\Rc_{k,\Sc_{+\!i}}^{\Sc}\!(\Dc),
\end{align}
and the total size of these residual IVs in bits is
\begin{align}\label{size_TildeR}
|\Upsilon_{k,\Sc}^{\Dc}|=TQ\sum_{i\in\Kc\backslash\Sc}r_{k,\Sc_{+\!i}}^{\Sc}\!(\Dc).
\end{align}

\subsubsection{Step 2: Nested Grouping of IVs}\label{sec:2step}

{Based on Step 1, in subset $\Sc$, we group the IVs needed by worker $k$ based on which worker subset they will be considered for coding. Let $\Ic_{k,\Sc}^\Dc$ denote the set of  IVs to be sent to worker $k$ within $\Sc$. It consists of the IVs in $\Vc^\Dc_{k,\Sc_{-\!k}}$ and all residual IVs in $\Upsilon_{k,\Sc}^{\Dc}$ that are from all $\Sc_{+i}$'s, for $i\in\Kc\backslash\Sc$,} \ie
\begin{align}\label{equ:I_CDC}
\Ic_{k,\Sc}^\Dc=\Vc^\Dc_{k,\Sc_{-\!k}}\bigcup \Upsilon_{k,\Sc}^{\Dc}.
\end{align}

From~\eqref{equ:size_V} and~\eqref{size_TildeR}, we have  the size of  $\Ic_{k,\Sc}^\Dc$ in bits as
\begin{align}\label{I_size}
|\Ic_{k,\Sc}^\Dc| = TQ\big(W_k\!\sum_{n\in\Dc}\!t_{n,\Sc_{-\!k}}+\!\sum_{i\in\Kc\backslash\Sc}r_{k,\Sc_{+\!i}}^{\Sc}\!(\Dc)\big).
\end{align}
Note that $|\Ic_{k,\Sc}^\Dc|$ is the total IVs needed by worker $k$ within $\Sc$, which may be different for each worker $k\in\Sc$.

{Let $\Ic^{\Dc}_{k,\Sc|j}$ denote the portion of $\Ic_{k,\Sc}^\Dc$  that are locally computed by worker $j\in\Sc\backslash\{k\}$, and to be sent to worker $k$ in $\Sc$.\footnote{In our notation, subscript $k$ indicates the worker who needs the IVs, and subscript $j$ represents the worker who sends the IVs.} We require all  $\Ic^{\Dc}_{k,\Sc|j}$'s, for  $k\in\Sc$, to be the same size, such that  worker $j$ can encodes them into one message without zero-padding and multicasts it to all other workers in $\Sc$ simultaneously. By this way, all IVs  are sent via multicasting. The size of this portion (sent to $\Sc$) in bits normalized by $TQ$ is denoted by $L_{j,\Sc}^\Dc\triangleq|\Ic^{\Dc}_{k,\Sc|j}|/TQ$,
which is the same for all other worker $k\in\Sc,k\neq j$.}

{After the above procedure, we remove $\Ic^{\Dc}_{k,\Sc|j}$ from $\Ic_{k,\Sc}^\Dc$, for all worker $j\in \Sc\backslash\{k\}$. If there are IVs left in  $\Ic_{k,\Sc}^\Dc$, we identify them as residual IVs to be coded in subset  $\Sc'\subset \Sc$ of size $|\Sc|-1$:}
{\begin{align}\label{equ:IVsub}
 \Ic_{k,\Sc}^\Dc -\sum_{j\in\Sc\backslash\{k\}}\!\!\Ic^{\Dc}_{k,\Sc|j}=\sum_{i\in\Sc\backslash\{k\}}\Rc_{k,\Sc}^{\Sc_{-\!i}}\!(\Dc)
 \end{align}}

{ Once all such subsets of the same size $|\Sc|$ are processed as the above, we will collect the residual IVs to be coded in a subset $\Sc_{-i}$ of $\Sc$, for each $i\in\Sc\backslash\{k\}$. The procedure essentially follows the same process of collecting the residual IVs as described in Step 1. In particular, let $\Sc^{\prime}=\Sc_{-i}$ for some $i\in\Sc, i\ne k$. The set of all residual IVs needed by worker $k$ that are collected from all the supersets of $\Sc^{\prime}$ of size $|\Sc^{\prime}|+1=|\Sc|$, \ie all $\Sc^{\prime}_{+j}$ for $j\in\Kc\backslash\Sc^{\prime}$. We note that $\Sc$ considered above is one of the supersets of $\Sc^{\prime}$. The residual IVs collected are from $\Sc$ as well as other supersets of size $|\Sc|$.}

{Note that in $\Ic_{k,\Sc}^\Dc$, those residual IVs that are  needed by worker $k$ in $\Sc$ and are encoded and sent to $\Sc_{-\!i}$ or subsets of $\Sc_{-\!i}$ is given by $\Rc_{k,\Sc}^{\Sc_{-\!i}}\!(\Dc)$, for $i\in\Sc\backslash\{k\}$.  It is of (normalized) size $r_{k,\Sc}^{\Sc_{-\!i}}\!(\Dc)$.
The total size of these residual IVs from $\Sc$  is $TQ\sum_{i\in\Sc\backslash\{k\}}r_{k,\Sc}^{\Sc_{-\!i}}\!(\Dc)$. Based on the above description on how the IVs in $\Ic_{k,\Sc}^\Dc$ are formed, the size of $\Ic_{k,\Sc}^\Dc$ can also be expressed as} 
{\begin{align}\label{I_size_CDC1}
|\Ic_{k,\Sc}^\Dc| =TQ\big(\sum_{j\in\Sc\backslash\{k\}}\!\!\!\!\!L_{j,\Sc}^{\Dc}\! +\sum_{i\in\Sc\backslash\{k\}}\!\!\!r_{k,\Sc}^{\Sc_{-\!i}}\!(\Dc)\big).
\end{align}}

The above nested coding process will continue until all IVs are encoded into messages for multicasting and no residual IVs remain.

We need to ensure that this nested coding strategy is feasible, \ie\ we can always regroup the residual IVs and send them to the subsets of the current worker subset. Thus, from~\eqref{I_size} and~\eqref{I_size_CDC1}, we have the following equality constraint on the size of $\Ic_{k,\Sc}^\Dc$:
\begin{align}\label{eq:con4}
&W_k\!\!\sum_{n\in\Dc}t_{n,\Sc_{-\!k}}\! \!+\!\!\!\sum_{i\in\Kc\backslash\!\Sc}\!\!\!r_{k,\Sc_{+i}}^{\Sc}\!(\Dc) =\!\!\!\! \sum_{j\in\Sc\backslash\{k\}}\!\!\!\!\!L_{j,\Sc}^{\Dc}\! +\!\!\!\!\sum_{i\in\Sc\backslash\{k\}}\!\!\!r_{k,\Sc}^{\Sc_{-\!i}}\!(\Dc),\nn\\
&\hspace{10em}k\in\Sc, \Sc\subseteq\Kc, |\Sc|\ge2, \Dc\subseteq\Nc.
\end{align}

Note that in this nested strategy for coded shuffling, $\{L_{j,\Sc}^{\Dc}\}$ and $\{r_{k,\Sc}^{\Sc_{-\!i}}\!(\Dc)\}$ are the design variables. To minimize the expected shuffling load, we need to jointly optimize these variables along  with the file placement $\Tc$. This CDC optimization problem will be described in Section~\ref{sec:placement}.

\subsubsection{Step 3: Coded Message Formation}\label{sec:3step}
After forming $\Ic^{\Dc}_{k,\Sc|j}$ for each $k\in\Sc\backslash\{j\}$, each worker $j \in \Sc$ generates a coded message via bitwise XOR operation on  $\Ic^{\Dc}_{k,\Sc|j}$'s, for all  $k\in\Sc\backslash\{j\}$, as
\vspace{-0.5em}
\begin{align}\label{equ:codedmsg}
C_{\Sc|j}^{\Dc} \triangleq \bigoplus_{k\in\Sc\backslash\{j\}}\Ic^{\Dc}_{k,\Sc|j},
\end{align}
which is then multicasted by worker $j$ to the rest workers in $\Sc$. 

To decode $C_{\Sc|j}^{\Dc}$ at worker $k \in \Sc$,  we note that by the definition of $\Ic^{\Dc}_{k,\Sc|j}$ in Step 2,  worker $k\in\Sc$ has already locally generated all the  IVs in $\Ic^{\Dc}_{i,\Sc|j}$'s, for $i\in\Sc\backslash\{ k, j\}$. Thus,  worker $k$ can successfully decode  $\Ic^{\Dc}_{k,\Sc|j}$ from  $C_{\Sc|j}^{\Dc}$  sent by worker $j\in \Sc\backslash\{k\}$. From~\eqref{equ:codedmsg}, the size of the coded message  $C_{\Sc|j}^{\Dc}$ in bits is
\begin{align}\label{equ:mgs_size}
|C_{\Sc|j}^{\Dc}|=TQL_{j,\Sc}^{\Dc}, \quad  j \in \Sc.
\end{align}


\subsubsection{{The Nested Coded Shuffling Procedure}}
{Our  nested coded shuffling strategy applies the three-step shuffling strategy described in Sections~\ref{sec:1step} --~\ref{sec:3step} to all the worker subsets in a top-down manner.  Starting from the entire worker set $\Kc$,  for each worker $k\in\Kc$, we have $\Ic_{k,\Kc}^\Dc=\Vc^\Dc_{k,\Kc_{-\!k}}$, since there are no residual IVs at this starting point.
Once the  three-step shuffling strategy is applied to $\Kc$, each worker $j\in\Kc, j\ne k$, will form a coded message to be multicasted to worker $k$ (and other workers). The residual IVs will be identified and distributed to the subsets containing worker $k$: $\{\Sc\subset\Kc, k\in\Sc,|\Sc|=K-1\}$.  For each of these subsets with size $K-1$, the same three-step procedure will be applied: 1) identify the residual IVs from all supersets of $\Sc$, and 2) form the coded messages to be multicasted to worker $k$ in $\Sc$. This nested procedure then continues  until $|\Sc|=1$ for subset $\Sc$ in consideration.
} 

\subsubsection{Example}
{ We illustrate the nested IV grouping and coding procedure in Steps 2 and 3  in Figs.~\ref{subfig1} and \ref{subfig2} using an example of three workers $\Kc=\{1,2,3\}$.
} 

{In Fig.~\ref{subfig1}, for  worker subset $\Sc=\Kc=\{1,2,3\}$, we describe how the IVs needed by each worker in $\Sc$ are partitioned and grouped into  coded messages for shuffling in $\Sc$ and identify the remaining IVs to be further considered for coding within the subsets of $\Sc$:  
}

\begin{itemize}[leftmargin=*]
\item 
{The left box shows the set $\Ic_{k,\Sc}^{\Dc}$ of the  IVs needed by each worker in $\Sc$, for $k=1,2,3$. Take $\Ic_{1,\Sc}^\Dc$ (needed by worker $1)$ as an example. It contains the needed IVs that are locally computed by workers 2 and 3. Only a fraction of $\Ic_{1,\Sc}^\Dc$ will be coded  by workers $2, 3$ for subset $\Sc$ and sent to worker $1$. 
}

\item { The middle box shows the coded message using $\Ic^{\Dc}_{k,\Sc|j}$'s by each worker in $\Sc$. For example, worker 2 has locally computed IVs  in $\Sc$ that are needed by the rest of workers 1 and 3 in $\Sc$.  For coding,  $\Ic^{\Dc}_{1,\Sc|2}$ (for worker 1) and $\Ic^{\Dc}_{3,\Sc|2}$ $ $(for worker 3) are of the same size and are coded into message $\Cc_{\Sc|2}^{\Dc}$ as in \eqref{equ:codedmsg}, which is multicasted to workers 1 and 3. Similarly, worker 3 forms   message $\Cc_{\Sc|3}^{\Dc}$ containing  $\Ic^{\Dc}_{1,\Sc|3}$. Thus, IVs in  $\Ic_{1,\Sc}^\Dc$  needed by worker 1 have been coded via  $\Ic^{\Dc}_{1,\Sc|2}$ and $\Ic^{\Dc}_{1,\Sc|3}$. 
}

\item {
 The right box shows the residual IVs from  $\Ic_{k,\Sc}^{\Dc}$ to be considered in different subsets of $\Sc$.
For  $\Ic_{1,\Sc}^\Dc$ needed by worker 1 in $\Sc$, the residual  IVs are $\Ic_{1,\Sc}^\Dc-\Ic^{\Dc}_{1,\Sc|2}-\Ic^{\Dc}_{1,\Sc|3}$ as  in (12).
They will be considered by workers 2, 3 for coding in the subsets of $\Sc$ that contain worker $1$, \ie $\Rc_{1,\Sc}^{\!\{1,2\}}\!(\Dc)$ in subset  $\{1,2\}$ at worker 2 and  $\Rc_{1,\Sc}^{\{1,3\}}(\Dc)$ in subset $\{1,3\}$ at worker 3. Different colors are used for different subsets  to help identify where the residual IVs will be grouped subsequently for coding. 
}
\end{itemize}

{In Fig.~\ref{subfig2}, we illustrate how the residual IVs in $\Rc_{k,\Sc}^{\!\Sc'}\!(\Dc)$'s from $\Sc$ are combined  in subset $\Sc'$ of  $\Sc$ for coding as in \eqref{equ:I_CDC}. Take subset $\{1,2\}$ as an example. Again,  $\Ic_{1,\{1,2\}}^{\Dc}$is the set of IVs needed by worker $1$ in $\{1,2\}$. It consists of 1) the IVs $\Vc_{1,\{2\}}^{\Dc}$ that are locally computed by worker 2 with the files  placed in  $\{2\}$  and 2) the residual IVs $\Rc_{1,\Sc}^{\!\{1,2\}}\!(\Dc)$ from $\Sc$ that are needed by worker $1$ and will be considered in $\{1,2\}$. Note that since $\Kc=\{1,2,3\}$, worker subset $\{1,2\}$ has only one superset $\Sc=\{1,2,3\}$. Thus, the residual IVs are only from $\Rc_{1,\Sc}^{\!\{1,2\}}\!(\Dc)$ in $\Sc$. For the general case, the residual IVs for  $\Ic_{1,\{1,2\}}^{\Dc}$ are collected from all supersets of $\{1,2\}$ as shown in (8), and $\Ic_{1,\{1,2\}}^{\Dc}$ is shown as in (10).
} 

\subsubsection{Validity of the Proposed Shuffling Strategy}
Our overall proposed nested coded shuffling strategy applies the above Steps 1 -- 3 to each worker subset in $\{\Sc:\Sc\subseteq\Kc, |\Sc|\ge2\}$, where  each worker in that subset generates the coded message to be multicasted to the other works in the same worker subset to complete the IV shuffling.  Note that a valid distributed computing scheme ensures that all the workers can obtain the IVs they need in the shuffling phase to compute the assigned target functions. We now show in the following proposition that along with the file placement, our proposed nested shuffling strategy is valid.

\begin{proposition}\label{proposition1}
The proposed file placement strategy and nested coded shuffling strategy for the heterogeneous CDC with nonuniform file popularity described in Section~\ref{sec:sys} are valid for any number of files $N$.
\end{proposition}
\IEEEproof
To show the validity, we need to show that for each job with its required set of files $\Dc$,   each worker $k\in\Kc$ can obtain all of the IVs it needs from $\Dc$. Recall that for each worker subset $\Sc$ where $|\Sc|\ge 2$, the  file placement strategy $\Tc$ and the shuffling  design variables $\{L_{k,\Sc}^{\Dc}, r_{k,\Sc}^{\Sc_{\!-\!i} }(\Dc)\}$ must satisfy \eqref{eq:con4}. We first show that for any given  $\Tc$, by the nested shuffling strategy in Steps $1$ to $3$, there is always a feasible solution 
$\{L_{k,\Sc}^{\Dc},r_{k,\Sc}^{\Sc_{-\!i}}(\Dc) \}$ satisfying~\eqref{eq:con4}.

To see this, for any given $\Sc$, the residual IVs needed by worker $k\in\Sc$ and computed by $\Sc_{-k}$ will be considered for coded shuffling in the smaller subsets of $\Sc$ of size $|\Sc|-1$. Furthermore, any residual IVs  in the subset of $\Sc$ of size $|\Sc|-1$ will be considered for the shuffling in its subsets of size $|\Sc|-2$.  Following this nested shuffling pattern, by the end of the procedure, any remaining portion of the residual IVs in $\Sc_{\!-\!k}$ that are needed by worker $k$ will be shuffled in the subsets of two workers $\{k,l\}$ for all $l\!\in\!\Sc_{\!-\!k}$. Note that for worker subset $\{k,l\}$ of size $2$, the IVs are unicasted to $k$ by worker $l$. This above procedure shows that the IVs needed by the workers in $\Sc$ can always be shuffled through the nested shuffling strategy.

Now we prove that each worker $k$ can receive all the needed IVs in the Shuffling phase. Summing both sides of~\eqref{eq:con4} over all  $\Sc$'s that have at least two workers and contain worker $k$, \ie $k\in\Sc$, and $\Sc\subseteq\Kc, |\Sc|\ge2$, we have
\begin{align}\label{eq:propo1}
&\sum_{\substack{\Sc\subseteq\Kc,\\ |\Sc|\ge2, k\in\Sc}}\!\!\!\!\!W_k\!\sum_{n\in\Dc}\!\!t_{n,\Sc_{-\!k}}\!+\!\sum_{\substack{\Sc\subseteq\Kc,\\ |\Sc|\ge2,k\in\Sc}}\sum_{i\in\Kc\backslash\!\Sc}\!\!\!r_{k,\Sc_{+i}}^{\Sc}\!(\Dc) =\nn\\
&\hspace{3em} \sum_{\substack{\Sc\subseteq\Kc,\\ |\Sc|\ge2,k\in\Sc}}\sum_{j\in\Sc\backslash\{ k\}}\!\!\!\!\!L_{j,\Sc}^{\Dc}\! +\!\sum_{\substack{\Sc\subseteq\Kc,\\ |\Sc|\ge3,k\in\Sc}}\sum_{i\in\Sc\backslash\{ k\}}\!\!\!r_{k,\Sc}^{\Sc_{-\!i}}\!(\Dc).
\end{align}

{The second terms on both sides of~\eqref{eq:propo1} are equivalent. To see this, note that at the left-hand side (LHS), the outer summation in the second term is for all subsets $\Sc\subseteq\Kc$ of size $|\Sc|\ge2$. For each $\Sc$, the inner summation contains all supersets of $\Sc$ with size $|\Sc|+1$ (\ie $\Sc_{+i}$, for $i\in \Kc\backslash \Sc$). 
At the right-hand side (RHS), the outer summation in the second term is for all $\Sc\subseteq\Kc$ of $|\Sc|\ge 3$. For each $\Sc$, the inner summation contains all subsets of $\Sc$ with size $|\Sc|-1$ (\ie  $\Sc_{-i}$, for $i\in\Sc\backslash\{k\}$).
We  see that the two terms at LHS and RHS are essentially two different ways (upward or downward) of counting each $\Sc\subseteq\Kc$ and all of its supersets of size $|\Sc|+1$, for $|\Sc|\ge 2$. Following this, we have}
\begin{align}\label{eq:propo2}
\sum_{\substack{\Sc\subseteq\Kc,\\ |\Sc|\ge2,k\in\Sc}}\!\!\!\!\!W_k\!\sum_{n\in\Dc}t_{n,\Sc_{-\!k}}=\sum_{\substack{\Sc\subseteq\Kc,\\ |\Sc|\ge2,k\in\Sc}}\sum_{j\in\Sc\backslash\{ k\}}\!\!\!\!\!L_{j,\Sc}^{\Dc}.
\end{align}
Note that the LHS of \eqref{eq:propo2} is the total size of all the IVs \emph{needed by} worker $k$ in all those worker subsets in $\{\Sc:\Sc\subseteq\Kc, |\Sc|\ge2, k\in\Sc\}$. As discussed at the beginning of Section~\ref{sec:shuffling}, this represents the size of all the IVs worker $k$ needs from other workers.  The RHS of~\eqref{eq:propo2} is the  total size of all the IVs \emph{sent by} all the workers other than worker $k$ in each subset in $\{\Sc:\Sc\subseteq\Kc, |\Sc|\ge2,k\in\Sc\}$. Thus, equality \eqref{eq:propo2} shows that, under our strategy satisfying constraint ~\eqref{eq:con4},  each worker $k$ can obtain its needed IVs from the received coded messages sent by all the other workers.
\endIEEEproof

\section{Optimization Of the Heterogeneous CDC}\label{sec:placement}

In this section, we  jointly optimize the file placement strategy in Section~\ref{sec:mapping} and the nested coded shuffling strategy   in Section~\ref{sec:shuffling} to minimize the expected shuffling load of the heterogeneous CDC.

For a given job with the required set of files $\Dc$, the shuffling load within a worker subset $\Sc$ is equal to the size of coded messages multicasted by all the workers in $\Sc$,  \ie $\sum_{j\in\Sc} |C_{\Sc|j}^{\Dc}|$. Thus, the overall shuffling load for this job, denoted by  $L(\Dc)$, is the sum of the shuffling loads in all the worker subsets in  $\{\Sc:\Sc\subseteq\Kc, |\Sc|\ge2\}$:
\begin{align}\label{equ:L_CDC}
\!L(\Dc)\!\!=\sum_{\Sc\subseteq\Kc,|\Sc|\ge 2}\ \sum_{j\in\Sc} |C_{\Sc|j}^{\Dc}|=\!\!\sum_{\Sc\subseteq\Kc,|\Sc|\ge 2}\ \sum_{j\in\Sc} L_{j,\Sc}^{\Dc}.
\end{align}
Let $p_{\Dc}$ denote the probability of the set of files $\Dc$  being accessed by a job. Note that $p_{\Dc}$ is a function of the file popularity distribution $\pbf$.
 Then, the  expected shuffling load is given by
\begin{align}\label{equ:avg_L}
\bar L=\sum_{\Dc\subseteq \Nc, \Dc \neq \emptyset}p_{\Dc}L(\Dc).
\end{align}
Our goal is to   jointly optimize the file placement  $\Tc$  and the nested coded shuffling strategy $\{L_{k,\Sc}^{\Dc}, r_{k,\Sc}^{\Sc_{\!-\!i} }(\Dc)\}$ to minimize the expected shuffling load   $\bar L$. With the file placement constraints~\eqref{eq:con1}--\eqref{eq:con3} and the feasibility constraint in \eqref{eq:con4} for the  nested shuffling, this problem is formulated as
\begin{align}
\ \hspace{-1.2em}\textrm{\bf P0}\!:\!\!\!&\min_{\{t_{n\!,\Sc}, L_{k,\Sc}^{\Dc}, r_{k,\Sc}^{\Sc_{\!-\!i}}(\Dc) \}}\; \bar L  \nn\\\textrm{s.t.} &\;\;\eqref{eq:con1}, \eqref{eq:con2}, \eqref{eq:con3}, \eqref{eq:con4},\ \text{and} \nn\\
&\hspace{.5em}L_{k,\Sc}^{\Dc}\ge 0,\quad k\in\Sc, \Sc\subseteq\Kc, |\Sc|\ge2, \Dc\subseteq\Nc, \label{equ:cons_ge0}\\
&\hspace{.5em}r_{k,\Sc}^{\Sc_{\!-\!i}}\!(\Dc) \!\!\ge\!0, \ k\!\in\!\Sc, i\!\in\! \Sc_{-\!k},\Sc\!\subseteq\!\Kc, |\Sc|\!\ge\!3,\Dc\!\subseteq\!\Nc.\ \label{equ:cons_ge1}
\end{align}

Note that since $t_{n,\Sc}\in\{0,1\}$,   {\bf P0} is an MILP problem, which is   NP-hard in general~\cite{conforti2014integer}.
For such an MILP problem, we can use the existing optimization solvers (\eg MOSEK~\cite{Weise:MOSEK21}, Gurobi etc.) to  obtain an approximate solution via the branch-and-cut method~\cite{tahernejad2020branch}. However, the computational complexity for this method is typically very high. To overcome this issue, in the following, we  develop a heuristic low-complexity  algorithm to  find a suboptimal solution to {\bf P0}.

\subsection{A Low-Complexity File-Group-Based Scheme}\label{sec:appro}

Instead of solving {\bf P0} for the optimal file placement indicators $\{t_{n\!,\Sc}\}$ in {\bf P0}, we first propose a two-file-group-based file placement strategy.

\emph{Two-file-group-based file placement:}   We partition the files in $\Nc$ into two non-overlapping groups. The first group contains $N_1$  most popular files in $\Nc$, denoted by $\Nc_1\triangleq\{1,\ldots,N_1\}$. The second group contains the rest of the less popular files, denoted by $\Nc_2\triangleq\Nc\backslash\Nc_1$.
The files in these two groups are placed at workers as follows:
\begin{itemize}
\item We start with the placement of files in the second file group $\Nc_2$ first. We place each file in $\Nc_2$  at exactly one worker in a round-robin fashion. Specifically, starting from $k=1$ and $n=N_1+1$, we place file $n$ at worker $k$ and update its mapping load $M_k$.
Then, we  place  the next file $n+1$ at the next  worker $(k+1) \bmod K$  until all files in $\Nc_2$ are placed at exactly one worker. 

\item For the  first group $\Nc_1$,
starting from worker $k=1$ and file $n=1$,  we place the files  from $\Nc_1$ at worker $k$ sequentially until the worker's mapping load $M_k$ is reached. Then,   we move to the next  worker and  repeat the same process by placing the subsequent files   from $\Nc_1$ at the worker.
Note that the files in $\Nc_1$ are selected  to be placed at workers in a cyclical manner until all the workers' mapping loads are met. 
\end{itemize}
The two-file-group-based placement strategy is described in Algorithm~\ref{alg:placement}. 
 For given $N_1$, Algorithm~\ref{alg:placement} computes the the set of files to be placed at worker $k$, denoted by
 $\overline\Mc_k(N_1)$, for all $k\in\Kc$. 
\begin{algorithm}[t]
\caption{Two-file-group-based file placement algorithm}\label{alg:placement}
\begin{algorithmic}[1]
\Require
   { $K$,  $M$,  $N$, $N_1$ }
\Ensure {$\overline\Mc_1(N_1),\ldots,\overline\Mc_K(N_1)$}

\hspace{-2em}$//$ File placement for the second file group $\Nc_2$
\State $\overline\Mc_k(N_1)=\emptyset$, $k \leftarrow 1$
\For {$n=N_1+1$ to $N$}
\If{$M_k >0$}
        \State $\overline\Mc_k(N_1) \leftarrow \overline\Mc_k(N_1)\cup\{n\}$
        \State $M_k \leftarrow M_k-1$
\EndIf

\State $k \leftarrow (k+1) \bmod K$
\EndFor

\hspace{-2em}$//$ File placement for the first file group $\Nc_1$
\State $n \leftarrow 0$
\For {$k=1$ to $K$}
\If{$M_k >0$}
        \For {$m=1$ to $M_k$}
        \State $\overline\Mc_k(N_1) \leftarrow \overline\Mc_k(N_1)\cup\{{(n+m) \bmod N_1}\}$
        \EndFor
\State $n \leftarrow (M_k+n) \bmod N_1$
\EndIf
\EndFor
\end{algorithmic}
\end{algorithm}

\begin{Remark} 

Note that the shuffling in the CDC adopts the similar technique used for delivery in coded caching~\cite{Li&Maddah-Ali:TIT2018}.
In both problems, the coded multicasting opportunities are exploited for content delivery (IV shuffling) to reduce the delivery (shuffling) load. For coded caching under nonuniform file popularity, the two-file-group-based cache placement strategy has been proposed~\cite{Zhang&Coded:TIT18,Ji&Order:TIT17,Deng&Dong:TIT22,Deng&DongMCCS:TIT22}, in which less popular files are stored solely in the server, and only popular files are stored at local caches.  It has been widely used in the literature due to its simplicity and has been shown to perform close to the optimal placement~\cite{Deng&DongMCCS:TIT22}. Our two-file-group-based file placement for the heterogeneous CDC
is inspired by this scheme. It employs a similar idea: We separate files into a popular file group $\Nc_1$ and a less popular file group $\Nc_2$. We place each file in $\Nc_2$ at one worker only to reduce the mapping loads used for these less popular files to the minimum. We then use all the remaining mapping loads to store the popular files in $\Nc_1$.
\end{Remark}

We now solve {\bf P0} under the proposed two-file-group-based file placement strategy. Denote the two-file-group-based file placement strategy with file partition at $N_1$ as $T_\text{fg}(N_1)\triangleq\{ t_{n,\Sc}(N_1): n\in \Nc, \Sc\subseteq\Kc, \Sc \neq \emptyset\}$, where the file placement indicator $t_{n,\Sc}(N_1)$ is given by
\begin{align}\label{equ:2groups}
t_{n,\Sc}(N_1)\!=\!
\begin{cases}
1,\ \ \text{for~} \Sc=\{k:k\in\Kc,n\in\overline\Mc_k(N_1)\} \\
0,\ \ \text{otherwise.}
\end{cases}
\end{align} 
Let $\bar L (T_\text{fg}(N_1))$ denote the expected shuffling load under $T_\text{fg}(N_1)$. Then, under this two-file-based file placement $T_\text{fg}(N_1)$,  {\bf P0} is reduced  to a nested coded shuffling optimization problem for $\{L_{k,\Sc}^{\Dc}, r_{k,\Sc}^{\Sc_{\!-\!i}}(\Dc) \}$ to minimize the expected shuffling load $\bar L(T_\text{fg}(N_1))$: 
\begin{align}
&\textrm{\bf P1}(T_\text{fg}(N_1)):\min_{\{L_{k,\Sc}^{\Dc}, r_{k,\Sc}^{\Sc_{\!-\!i}}(\Dc) \}} \bar L (T_\text{fg}(N_1))  \nn\\ 
&\hspace{0em}\text{s.t.}\ \  \eqref{equ:cons_ge0}, \eqref{equ:cons_ge1}, \text{and}\nn\\
&W_k\!\!\sum_{n\in\Dc}t_{n,\Sc_{-\!k}}(N_1)\! \!+\!\!\!\sum_{i\in\Kc\backslash\!\Sc}\!\!\!r_{k,\Sc_{+i}}^{\Sc}\!(\Dc) =\!\!\!\! \sum_{j\in\Sc, j\neq k}\!\!\!\!\!L_{j,\Sc}^{\Dc}\! +\!\!\!\!\sum_{i\in\Sc, i \neq k}\!\!\!r_{k,\Sc}^{\Sc_{-\!i}}\!(\Dc),\nn\\
&\hspace{9.5em}k\in\Sc, \Sc\subseteq\Kc, |\Sc|\ge2, \Dc\subseteq\Nc.\label{equ:P1_cons3}
\end{align}

Note that {\bf P1}($T_{\text{fg}}(N_1)$) is an LP problem and we can apply  the standard LP solvers to compute the solution. According to~\eqref{equ:cons_ge0}, the total number of   $L_{k,\Sc}^{\Dc}$'s is $\left(2^N-1\right)\sum_{k=2}^{K}\binom{K}{k}k$, and in~\eqref{equ:cons_ge1}, the total number of $r_{k,\Sc}^{\Sc_{\!-\!i}}(\Dc)$'s is $\left(2^N-1\right)\sum_{k=3}^{K}\binom{K}{k}k^2$. Thus,  the total number of variables in {\bf P1}($T_{\text{fg}}(N_1)$) is $\left(2^N-1\right)\!\left( K^2+\sum_{k=3}^{K}\binom{K}{k}k^2-K\right)$. Further, from~\eqref{equ:cons_ge0}, \eqref{equ:cons_ge1}, and \eqref{equ:P1_cons3},  {\bf P1}($T_{\text{fg}}(N_1)$) contains a total number of $\left(2^N-1\right)\!\left(
2K^2+\sum_{k=3}^{K}\binom{K}{k}(k^2+k)-2K\right)$ constraints. 

\begin{algorithm}[t]
\caption{Low-complexity  two-file-group-based approximate solution for {\bf P0}}\label{alg:p0}
\begin{algorithmic}[1]
\Require
   { $K$,  $M$,  $N$, $N_1$. }
\Ensure {$T_\text{fg}(N_1^*)$, $\bar L (T_\text{fg}(N_1^{*}))$.}
\For {$N_1=1$ to $N$}
\State  Compute the two-file-group-based file placement strategy $\overline\Mc_k(N_1)$ and  $T_\text{fg}(N_1)$ by Algorithm~\ref{alg:placement} and~\eqref{equ:2groups}, respectively.
\State Under $T_\text{fg}(N_1)$, solve {\bf P1} to obtain the minimum $\bar L (T_\text{fg}(N_1))$ for given $T_\text{fg}(N_1)$. \EndFor
\State Compute $N_1^*=\argmin_{N_1\in N}\bar L (T_\text{fg}(N_1))$.
\State Return $T_\text{fg}(N_1^*)$, $\bar L (T_\text{fg}(N_1^{*}))$.  
\end{algorithmic}
\end{algorithm}

Given the two-file-group-based placement strategy $T_\text{fg}(N_1)$ for $N_1\in\Nc$, {\bf P1}($T_\text{fg}(N_1)$) optimizes the coded shuffling strategy. Following the above, we can further optimize the two-file-group-based placement strategy by searching for the optimal $N_1^*\in\Nc$ that leads to the minimum expected shuffling load $\bar L (T_\text{fg}(N_1^*))$. We summarize the overall algorithm in Algorithm~\ref{alg:p0}. {In Section~\ref{sec:simu}, we will show through the numerical results that  the performance of  Algorithm~\ref{alg:p0} is very close to that of solving {\bf P0} directly via the branch-and-cut method, but with significantly less computational complexity.} 

{We have proposed a two-file-group-based file placement algorithm. Increasing the number of file groups will significantly increase the design complexity of the corresponding file placement algorithm. In principle, there could be as many as  $N$ file groups if we treat every file differently for the placement. However, as it will be shown in our numerical study that the performance of Algorithm~\ref{alg:p0} is very close to that of solving {\bf P0} directly, it implies that considering such a more complicated multi-group file placement strategy would not be very beneficial for reducing shuffling load.}

{
\begin{Remark}
Note that the proposed two-file-group-based file placement algorithm is different from the two-file-group-based \textit{cache} placement proposed for coded caching~\cite{Zhang&Coded:TIT18,Ji&Order:TIT17,Deng&Dong:TIT22,Deng&DongMCCS:TIT22}. Firstly, the cache placement for coded caching needs to partition each file into subfiles. In contrast, for the CDC, the entire file is stored at a worker  without partitioning. Secondly, the files may not be cached by any user (only stored by the server) in coded caching. In particular, the two-file-group-based cache placement assigns zero cache to the files in the less popular group. However, for the CDC, each file must be stored by at least one worker.  Thus, the two-file-group-based cache placement cannot be applied to the CDC. Instead, our proposed two-file-group-based algorithm employs a round-robin strategy to place the files in the less popular group, which has not been considered in existing works.
\end{Remark}
}

\section{Heterogeneous Compressed Coded Distributed Computing}\label{sec:C-CDC}
In this section, we consider applying a data compression technique to further reduce the shuffling load. In particular, we propose a heterogeneous compressed CDC (C-CDC) scheme, where a {lossless} data compression technique is used along with the nested coded shuffling strategy proposed in Section~\ref{sec:shuffling}. The heterogeneous C-CDC is particularly useful for the class of computing jobs, where the target functions are linear aggregation functions.   

\subsection{Computing Jobs with Aggregate Target Functions}\label{sec:C-CDC1}

We first describe a type of computing jobs, whose target functions are linear aggregation functions. In particular, for a job with files $\Dc$, the target function $\phi_q(\Dc)$ in~\eqref{equ:reduce_func} is the aggregation of the IVs in $\{V_{q,n}, n\in\Dc\}$, given by
\begin{align}\label{equ:aggre_IV}
\phi_q(\Dc)=h_q(\{V_{q,n}, n\in\Dc\})=\sum_{n\in\Dc}V_{q,n},\quad  q\in\Qc.
\end{align}

\emph{Example:} Computing jobs with aggregate target functions are commonly seen in distributed training for machine learning problems. For example, in a typical machine learning problem, the goal is to learn a global model represented by the global model parameter vector $\thetabf=[\thetabf_1^T,\ldots,\thetabf_Q^T]^T$ based on the input data sample (file) set $\Dc$. The learning objective is  to obtain the optimal global model $\thetabf^*$ that minimizes the sum of the loss functions over all the files $\Dc$, given by
\begin{align}\label{equ:LR}
\thetabf^*=\argmin_{\thetabf}\sum_{n\in\Dc}J(\thetabf,n)+\lambda R(\thetabf),
\end{align} 
where $J(\thetabf,n)$ is the loss function for file $n\in\Dc$, $\lambda>0$ is the regularization parameter, and $R(\thetabf)$ is the regularization function. The learning problem described in~\eqref{equ:LR} can be solved by the gradient descent algorithm. Let $\thetabf^{t}$ denote the current model parameter computed at iteration $t$. At  iteration $t+1$, the global model update for $\thetabf^t_q$ is given by
\begin{align}\label{equ:CCDC_update}
\!\!\thetabf^{t+1}_q=\thetabf^{t}_q-\alpha\sum_{n\in\Dc}\nabla_{\thetabf_q} J(\thetabf^t,n)-\lambda\nabla_{\thetabf_q} R(\thetabf^t),\   q\in\Qc,
\end{align}
where $\alpha$ is the learning rate.

Now we consider computing the update in~\eqref{equ:CCDC_update} distributedly using the MapReduce framework.  In this case, each IV $V_{q,n}$ can be viewed as the gradient of the loss function with respect to (w.r.t.) $\thetabf_q$ and file $n$, given by 
\begin{align}
V_{q,n}=\nabla_{\thetabf_q} J(\thetabf^{t},n),\ \ \text{for } q\in\Qc, n\in\Dc .
\end{align}The target function $\phi_q(\Dc)$ is given by 
\begin{align}
\phi_q(\Dc)=\sum_{n\in\Dc}\nabla_{\thetabf_q} J(\thetabf^{t},n)
\end{align} 
which is the aggregation of the gradients for the set of files $\Dc$ needed for the model update in~\eqref{equ:CCDC_update}.
Each worker $k$ is assigned a subset of target functions $\{\phi_q^{t+1}(\Dc),q\in\Wc_k\}$, which is essentially to compute the global model updates $\{\thetabf_q^{(t+1)},q\in\Wc_k\}$. 

\subsection{Heterogeneous C-CDC}

We now propose a heterogeneous C-CDC scheme for the jobs that have aggregate target functions as shown in~\eqref{equ:aggre_IV}. We again consider a general scenario where the file popularities and users' mapping and reducing loads are all nonuniform. 

The key idea of the C-CDC is to let  each worker aggregates its local IVs before the nested coded shuffling phase. Recall that for the CDC described in Section~\ref{sec:problem}, for worker $k$ in $\Sc\subseteq\Kc$, it needs to obtain  the IVs  computed by all the  workers in $\Sc_{-\!k}$,  \ie $\Vc^\Dc_{k,\Sc_{-\!k}}=\{V_{q,n}:q\in\Wc_k, n\in \Ac_{\Sc_{-\!k}}^\Dc \}$. For the C-CDC, we can rewrite the target function assigned to worker $k$, $\phi_q(\Dc), q\in\Wc_k$, into the following equivalent form:
\begin{align}\label{equ:target_agg}
\!\!\!\phi_q(\Dc)=\!\!\!\sum_{n\in \Mc_k\cap\Dc}V_{q,n}+\!\!\!\sum_{\Sc:\Sc\subseteq\Kc,k\in\Sc}\sum_{n\in \Ac_{\Sc_{-\!k}}^\Dc}V_{q,n}, \ q\in\Wc_k.\!
\end{align}
{The first term  on the RHS of~\eqref{equ:target_agg} is the aggregation of IVs for the files in $\Mc_k\cap\Dc$ that are placed at worker $k$. The second term on the RHS of~\eqref{equ:target_agg} is the aggregation of IVs for the files that are not placed at worker $k$. In particular, it is computed by  obtaining the aggregated IVs  for the files placed in $\Sc_{-k}$, \ie $\sum_{n\in \Ac_{\Sc_{-\!k}}^\Dc}V_{q,n}$,   and then summing them over all $\Sc$'s that contain worker $k$.} Note that the second term on the RHS of~\eqref{equ:target_agg} represents the aggregated IVs that are needed by worker $k$. 

Based on the above, in the shuffling phase, each worker $k$ in $\Sc$ needs to obtain the \emph{aggregated} IVs computed by workers in $\Sc_{-\!k}$, for each $q\in\Wc_{k}$, given by 
\begin{align}\label{equ:C-CDC_IV}
\widetilde\Vc^\Dc_{k,\Sc_{-\!k}}=\bigg\{\sum_{ n\in \Ac_{\Sc_{-\!k}}^\Dc }V_{q,n}:q\in\Wc_k\bigg\}
\end{align}
Compare with~\eqref{equ:IV_in_S}, we see that the set of IVs needed by worker $k$ is now compressed through aggregation as in~\eqref{equ:C-CDC_IV}, which is to be used for the nested coded shuffling. Thus, the size of the contents to be shuffled, \ie $\widetilde\Vc^\Dc_{k,\Sc_{-\!k}}$, is further reduced in the C-CDC as compared with that of the CDC in~\eqref{equ:IV_in_S}. {Note that the aggregated IVs will be decoded and used by worker $k$ directly to compute its assigned target function $\phi_q(\Dc)$ without the need for decompression. Thus, the C-CDC does not lose computing accuracy for jobs with aggregate target functions. We can view  the C-CDC is an improved version that removes some redundancy in the original CDC.}

\subsubsection{The Mapping Strategy}For the heterogeneous C-CDC, we follow the same model of file placement strategy described in Section~\ref{sec:mapping} for  the heterogeneous CDC. Specifically, in the Map phase, we again use the file placement indicator $t_{n,\Sc}$ to indicate whether file $n$ is stored exclusively at worker subset  $\Sc\subseteq\Kc$. As a result, we have the file placement constraints on $\{t_{n,\Sc}\}$ in~\eqref{eq:con1}--\eqref{eq:con3}.

\subsubsection{The Nested Coded Shuffling with IV Aggregation}
The key idea of the C-CDC is to add an extra step of IV aggregation in the nested coded shuffling strategy described in Section~\ref{sec:shuffling}. In general, the shuffling strategy of the C-CDC follows the three-step of nested coded shuffling  for the CDC described in Sections~\ref{sec:1step} -- \ref{sec:3step}, with some modifications as described below. 

\textbf{Step 1:} For worker $k\in\Sc$, we first determine the aggregated IVs  needed by worker $k$ in $\Sc_{-k}$. To compute target function $\phi_q(\Dc), q\in\Wc_k$, worker $k$ needs aggregated IVs $\widetilde\Vc^\Dc_{k,\Sc_{-\!k}}$ in~\eqref{equ:C-CDC_IV}. 
The size of $\widetilde\Vc^\Dc_{k,\Sc_{-\!k}}$ depends on $\Ac_{\Sc_{-\!k}}^\Dc$, \ie whether there are files exclusively placed in $\Sc_{-\!k}$. If $\Ac^{\Dc}_{\Sc_{-\!k}}=\emptyset$, no IVs from $\Sc_{-k}$ needed for shuffling and $|\sum_{ n\in \Ac_{\Sc_{-\!k}}^\Dc }V_{q,n}|=0$, for $q\in\Wc_k$. Otherwise, its size is equal to the size of a single IV: $|\sum_{ n\in \Ac_{\Sc_{-\!k}}^\Dc }V_{q,n}|=T$. Following this, we have
\begin{align}\label{equ:IVinS_size}
|\widetilde\Vc^\Dc_{k,\Sc_{-\!k}}|=
\begin{cases}
0,\quad &\text{if}\ \Ac^{\Dc}_{\Sc_{-\!k}}=\emptyset\\
W_kTQ,\quad &\text{otherwise},
\end{cases}
\end{align}
which can be further expressed as a function of file placement indicator $t_{n,\Sc_{-\!k}}$ as 
\begin{align}
|\widetilde\Vc^\Dc_{k,\Sc_{-\!k}}|=W_kTQ\max_{n\in\Dc}t_{n,\Sc_{-\!k}}.
\end{align}

{As discussed in Section~\ref{sec:1step}, for each worker $j\in\Sc, j\ne k$, only a fraction of  aggregated IVs needed by worker $k$ will be coded by worker $j$ and sent to  worker $k$. Similar to the definition of residual IVs, we define  $\widetilde \Rc_{k,\Sc_{+\!i}}^{\Sc}\!(\Dc)$
as the \emph{residual aggregated IVs}, which is a portion of the aggregated IVs needed by worker $k$ in $\Sc_{+i}$, but will be coded
in the subsets of $\Sc$. Also, define $\tilde r_{k,\Sc_{+\!i}}^{\Sc}\!(\Dc)\triangleq|\widetilde\Rc_{k,\Sc_{+\!i}}^{\Sc}\!(\Dc)|/TQ$
as the size of $\widetilde\Rc_{k,\Sc_{+\!i}}^{\Sc}\!(\Dc)$ in bits normalized
by $TQ$.}

For the delivery of the aggregated IVs and residual aggregated IVs, we follow the same nested coded shuffling procedure in  Steps 2 and 3 described in Sections~\ref{sec:2step} and~\ref{sec:3step}, respectively. The only difference is that instead of the  original IVs $\Vc^\Dc_{k,\Sc_{-\!k}}$ and residual IVs $ \Rc_{k,\Sc_{+\!i}}^{\Sc}\!(\Dc)$, we now consider shuffling the aggregated IVs $\widetilde\Vc^\Dc_{k,\Sc_{-\!k}}$ and residual aggregated IVs $\widetilde \Rc_{k,\Sc_{+\!i}}^{\Sc}\!(\Dc)$, for any worker $k\in\Sc$. 

\textbf{Step 2:} Similar to~\eqref{equ:I_CDC}, we group all the aggregated IVs needed by worker $k$ in $\Sc$ to one set, denoted by $\widetilde\Ic_{k,\Sc}^\Dc$, given by
\begin{align}
\widetilde\Ic_{k,\Sc}^\Dc=\widetilde\Vc^\Dc_{k,\Sc_{-\!k}}\bigcup \left(\bigcup_{i\in\Kc\backslash\Sc}\widetilde\Rc_{k,\Sc_{+\!i}}^{\Sc}\!(\Dc)\right),
\end{align}
which consists of the aggregated IVs in $\widetilde\Vc^\Dc_{k,\Sc_{-\!k}}$ and all residual aggregated IVs in $\widetilde\Rc_{k,\Sc_{+\!i}}^{\Sc}\!(\Dc)$ that are from all $\Sc_{+i}$'s, for $i\in\Kc\backslash\Sc$. The size of $\widetilde\Ic_{k,\Sc}^\Dc$ is given by
\begin{align}\label{I_size1}
|\widetilde\Ic_{k,\Sc}^\Dc| = W_kTQ\max_{n\in\Dc}t_{n,\Sc_{-\!k}}+TQ\!\sum_{i\in\Kc\backslash\Sc} \tilde r_{k,\Sc_{+\!i}}^{\Sc}\!(\Dc).
\end{align}


Following the same process of the nested grouping of IVs for the CDC in Section~\ref{sec:2step}, we consider the content in 
$\widetilde\Ic_{k,\Sc}^\Dc$ that are either coded in $\Sc$ or considered as residual aggregated IVs, and repartition it for nested coded shuffling. We first let $\widetilde\Ic^{\Dc}_{k,\Sc|j}$ denote the portion of $\widetilde\Ic_{k,\Sc}^\Dc$  that are locally computed by worker $j\in\Sc$ and to be sent to worker $k\in\Sc,k\ne j$. We again require  $\widetilde\Ic^{\Dc}_{k,\Sc|j}$'s, for  $k\in\Sc\backslash\{j\}$, to be the same size to avoid zero-padding. Following this, we denote the normalized size of $\widetilde\Ic^{\Dc}_{k,\Sc|j}$ by 
$\widetilde L_{j,\Sc}^\Dc\triangleq|\widetilde\Ic^\Dc_{k,\Sc|j}|/TQ$, which is the same for all $k\in\Sc,k\neq j$. 
Furthermore, we identify the residual aggregated IVs remained in  $\widetilde\Ic_{k,\Sc}^\Dc$, which will be coded in subset  $\Sc'\subset \Sc$ of size $|\Sc|-1$. According to Step 1,
the set of residual aggregated IVs to be encoded
and sent to $\Sc_{-\!i}$ or subsets of $\Sc_{-\!i}$, for $i\in\Sc\backslash\{k\}$
is of (normalized) size $\tilde r_{k,\Sc}^{\Sc_{-\!i}}\!(\Dc)$.
Thus, the total size of these residual aggregated IVs from $\Sc$  is $TQ\sum_{i\in\Sc\backslash\{k\}}\tilde r_{k,\Sc}^{\Sc_{-\!i}}\!(\Dc)$. {Note that after removing all $\widetilde \Ic^{\Dc}_{k,\Sc|j}$'s that are coded by the workers in $\Sc\backslash\{k\}$  from $\widetilde\Ic_{k,\Sc}^\Dc$, we have the residual aggregated IVs from $\widetilde\Ic_{k,\Sc}^\Dc$: }
{\begin{align}\label{equ:IVsub_C}
\widetilde\Ic_{k,\Sc}^\Dc -\sum_{j\in\Sc\backslash\{k\}}\!\!\widetilde\Ic^{\Dc}_{k,\Sc|j}=\sum_{i\in\Sc\backslash\{k\}}\widetilde\Rc_{k,\Sc}^{\Sc_{-\!i}}\!(\Dc) \end{align}}
As a result, the size of $\widetilde \Ic_{k,\Sc}^\Dc$ can also be expressed as 
\begin{align}\label{I_size_C-CDC1}
|\widetilde\Ic_{k,\Sc}^\Dc| =TQ\big(\sum_{j\in\Sc\backslash\{k\}}\!\!\widetilde L_{j,\Sc}^{\Dc}\!
+\sum_{i\in\Sc\backslash\{k\}}\!\tilde r_{k,\Sc}^{\Sc_{-\!i}}\!(\Dc)\big).
\end{align}

Based on \eqref{I_size1} and \eqref{I_size_C-CDC1}, to ensure that the nested coding strategy for the heterogeneous C-CDC is feasible, similar to \eqref{eq:con4},  we form the nested coded shuffling constraint as
\begin{align}\label{eq:con4_compressed}
&\max_{n\in\Dc}t_{n,\Sc_{-\!k}}W_k+\!\!\!\sum_{i\in\Kc\backslash\!\Sc}\!\!\!\tilde r_{k,\Sc_{+i}}^{\Sc}\!(\Dc) =\!\!\!\! \sum_{j\in\Sc, j\neq k}\!\!\!\!\!\widetilde L_{j,\Sc}^{\Dc}\! +\!\!\!\!\sum_{i\in\Sc, i \neq k}\! \tilde r_{k,\Sc}^{\Sc_{-\!i}}\!(\Dc),\nn\\
&\hspace{10em}k\in\Sc, \Sc\subseteq\Kc, |\Sc|\ge2, \Dc\subseteq\Nc.
\end{align}

\textbf{Step 3:} Each worker $j \in \Sc$ generates a coded message via bitwise XOR operation on  $\widetilde\Ic^{\Dc}_{k,\Sc|j}$'s, for all  $k\in\Sc\backslash\{j\}$, as
\vspace{-0.5em}
\begin{align}\label{equ:codedmsg1}
\widetilde C_{\Sc|j}^\Dc \triangleq \bigoplus_{k\in\Sc\backslash\{j\}}\widetilde\Ic^{\Dc}_{k,\Sc|j},
\end{align}
whose size
in bits is given by\begin{align}
|\widetilde C_{\Sc|j}^\Dc|=TQ\widetilde L_{j,\Sc}^\Dc,\quad j\in\Sc.
\end{align}

In the shuffling phase, the coded message $\widetilde C_{\Sc|j}^\Dc$ is multicasted by worker $j$ to the rest workers in $\Sc$.
Note that worker $k\in\Sc$ has already locally formed all the $\widetilde\Ic^{\Dc}_{i,\Sc|j}$'s, for $i\in\Sc\backslash\{ k, j\}$. Thus,  worker $k$ can successfully decode  $\widetilde\Ic^{\Dc}_{k,\Sc|j}$ from  $\widetilde C_{\Sc|j}^{\Dc}$  sent by worker $j\in \Sc\backslash\{k\}$. The overall shuffling strategy of the heterogeneous C-CDC applies Steps $1$ -- $3$ to each of the worker subset in $\{\Sc:\Sc\subseteq\Kc, |\Sc|\ge2\}$.

\begin{Remark}
{To the best of our knowledge, combining data compression and coded shuffling to reduce the shuffling load for executing jobs with aggregated target functions has only been studied in~\cite{Elkordy&Li2021}. However, the study conducted in~\cite{Elkordy&Li2021} is under a much simpler setup than the model in our work. In particular, it only considers a special case of homogeneous CDC, where all the files are needed by the job, and workers' mapping and reducing loads are equal. Under this homogeneous system setup, the IVs to be coded in the message are of equal size, and  no residual IV exists in the shuffling phase. Moreover, the scheme in~\cite{Elkordy&Li2021} is designed by assuming a sufficiently large number of files for its validity and thus is only applicable in such cases. These limitations greatly hinder the practical use of the CDC. In contrast, our work addresses all the above limitations. We consider a general heterogeneous scenario where the file popularity is nonuniform, and users' mapping and reducing loads are unequal. To tackle the design challenges, we adopted a
nested coded shuffling mechanism to form coded messages in a nested manner, which is very different from the shuffling strategy used in~\cite{Elkordy&Li2021}. Furthermore, our scheme can be applied to systems with an arbitrary number of files. Lastly, unlike the heuristic design used in~\cite{Elkordy&Li2021}, we provide an optimization framework to optimize the heterogeneous C-CDC w.r.t. file placement and nested coded
shuffling variables to minimize the shuffling load.~}  
\end{Remark}
 
\subsection{Optimization of the Heterogeneous C-CDC}\label{sec:opt_C-CDC}

We further consider optimizing the heterogeneous C-CDC by jointly optimizing its file placement and nested coded shuffling strategy.

Denote the overall shuffling load for a given job by  $\widetilde L(\Dc)$.  Similar to the shuffling load $L(\Dc)$ in \eqref{equ:L_CDC} for the CDC in Section~\ref{sec:placement}, we have:
\begin{align*}
\!\widetilde  L(\Dc)\!\!=\sum_{\Sc\subseteq\Kc,|\Sc|\ge 2}\ \sum_{j\in\Sc} |\widetilde  C_{\Sc}^j (\Dc)|=\!\!\sum_{\Sc\subseteq\Kc,|\Sc|\ge 2}\ \sum_{j\in\Sc} \widetilde L_{j,\Sc}^{\Dc},
\end{align*}
and the  expected shuffling load is 
\begin{align}\label{equ:avg_L_C-CDC}
\bar L_\text{\tiny C-CDC}=\sum_{\Dc\subseteq \Nc, \Dc \neq \emptyset}p_{\Dc}\widetilde L(\Dc).
\end{align}

To minimize $\bar L_\text{\tiny C-CDC}$ under the heterogeneous C-CDC, we formulate joint optimization of the file placement  $\Tc$  and the nested coded shuffling strategy $\{\widetilde L_{k,\Sc}^{\Dc}, \tilde r_{k,\Sc}^{\Sc_{\!-\!i} }(\Dc)\}$. Given the file placement constraints  \eqref{eq:con1}--\eqref{eq:con3} and the nested coded delivery constraint in \eqref{eq:con4_compressed}, we have the following optimization problem: 
\begin{align}
\ \hspace{-1.2em}\textrm{\bf P2}\!:\!\!\!&\min_{\{t_{n\!,\Sc}, \widetilde L_{k,\Sc}^{\Dc}, \tilde r_{k,\Sc}^{\Sc_{\!-\!i}}(\Dc) \}}\; \bar L_\text{\tiny C-CDC}  \nn\\\textrm{s.t.} &\;\;\eqref{eq:con1}, \eqref{eq:con2}, \eqref{eq:con3},\eqref{equ:cons_ge0},\eqref{equ:cons_ge1}, \text{and}\ \eqref{eq:con4_compressed}.\nn
\end{align} 

We rewrite the $\max$ operation in constraint~\eqref{eq:con4_compressed},  and transform {\bf P2} into the following equivalent form:
\begin{align}
\ \hspace{-1.2em}\textrm{\bf P3}\!:\!\!\!&\min_{\{t_{n\!,\Sc}, \widetilde L_{k,\Sc}^{\Dc},\tilde  r_{k,\Sc}^{\Sc_{\!-\!i}}(\Dc) \}}\; \bar L_\text{\tiny C-CDC}  \nn\\
&\textrm{s.t.} \;\;\eqref{eq:con1}, \eqref{eq:con2}, \eqref{eq:con3},\eqref{equ:cons_ge0},\eqref{equ:cons_ge1}, \text{and}\ \nn\\
&t_{n,\Sc_{-\!k}}W_k+\!\!\!\sum_{i\in\Kc\backslash\!\Sc}\!\!\!\tilde r_{k,\Sc_{+i}}^{\Sc}\!(\Dc) -\!\!\!\! \sum_{j\in\Sc, j\neq k}\!\!\!\!\!\widetilde L_{j,\Sc}^{\Dc}\! -\!\!\!\!\sum_{i\in\Sc, i \neq k}\!\!\!\tilde r_{k,\Sc}^{\Sc_{-\!i}}\!(\Dc)\le0,\nn\\
&\hspace{6em}k\in\Sc, \Sc\subseteq\Kc, |\Sc|\ge2, n\in\Dc,\Dc\subseteq\Nc.\label{equ:P3_con1} \end{align} 

Problem {\bf P3} is again an NP-hard MILP
problem, of which the approximate solution can be computed numerically by branch-and-cut method using existing optimization solvers. However, due to the high computational complexity of this method, we again explore the two-file-group-based algorithm proposed in Section~\ref{sec:appro} to solve {\bf P3}.
Specifically, for given file group partition $N_1\in\Nc$,   Algorithm~\ref{alg:placement} computes the set of files to be placed at worker $k$,  $\overline\Mc_k(N_1)$ for all $k\in\Kc$, and the file placement vector $\widetilde T_{\text{fg}}(N_1)$ is obtained by~\eqref{equ:2groups}. Under the
two-file-group-based file placement $\widetilde T_{\text{fg}}(N_1)$, {\bf P2} is reduced to a nested coded shuffling optimization problem for   $\{\widetilde L_{k,\Sc}^{\Dc}, \tilde r_{k,\Sc}^{\Sc_{\!-\!i}}(\Dc) \}$ to minimize the expected shuffling load $\bar L_\text{\tiny C-CDC}(\widetilde T_{\text{fg}}(N_1))$ as 
\begin{align}
\ \hspace{-1.2em}\textrm{\bf P4}(\widetilde T_{\text{fg}}(N_1))\!:\!\!\!&\min_{\{\widetilde L_{k,\Sc}^{\Dc},\tilde  r_{k,\Sc}^{\Sc_{\!-\!i}}(\Dc) \}}\; \bar L_\text{\tiny C-CDC}(\widetilde T_{\text{fg}}(N_1))  \nn\\
&\textrm{s.t.} \;\;\eqref{equ:cons_ge0},\eqref{equ:cons_ge1}, \text{and}\ \eqref{equ:P3_con1}. \nn
\end{align}
Since the objective and constraint functions are all linear w.r.t. $\{\widetilde
L_{k,\Sc}^{\Dc},\tilde  r_{k,\Sc}^{\Sc_{\!-\!i}}(\Dc)\}$, {\bf P4}$(\widetilde T_{\text{fg}}(N_1))$ is an LP problem and can be solved by standard LP solvers. Following this, similar to that of the CDC in Section~\ref{sec:appro}, we can find the optimal $N_1^*$ by a search in $\Nc$.  The overall algorithm is similar to Algorithm~\ref{alg:p0} and thus is omitted for brevity. We will demonstrate the effectiveness of this scheme in the numerical studies in Section~\ref{sec:simu}.

\begin{algorithm}[t]
\caption{{Low-complexity  two-file-group-based approximate solution for {\bf P3}}}\label{alg:p1}
\begin{algorithmic}[1]
\Require
   { $K$,  $M$,  $N$, $N_1$. }
\Ensure {$\widetilde T_\text{fg}(N_1^*)$, $\bar L\text{\tiny C-CDC} (\widetilde T_\text{fg}(N_1^{*}))$.}
\For {$N_1=1$ to $N$}
\State  Compute the two-file-group-based file placement strategy $\overline\Mc_k(N_1)$ and  $\widetilde T_\text{fg}(N_1)$ by Algorithm~\ref{alg:placement} and~\eqref{equ:2groups}, respectively.
\State Under $\widetilde T_\text{fg}(N_1)$, solve {\bf P4} to obtain the minimum $\bar L_\text{\tiny C-CDC} (\widetilde T_\text{fg}(N_1))$ for given $\widetilde T_\text{fg}(N_1)$. \EndFor
\State Compute $N_1^*=\argmin_{N_1\in N}\bar L\text{\tiny C-CDC} (\widetilde T_\text{fg}(N_1))$.
\State Return $\widetilde T_\text{fg}(N_1^*)$, $\bar L\text{\tiny C-CDC} (\widetilde T_\text{fg}(N_1^{*}))$.  
\end{algorithmic}
\end{algorithm}

\section{Numerical Results}\label{sec:simu}

\begin{figure}[t]
  \centering
  {\includegraphics[width=1\linewidth]{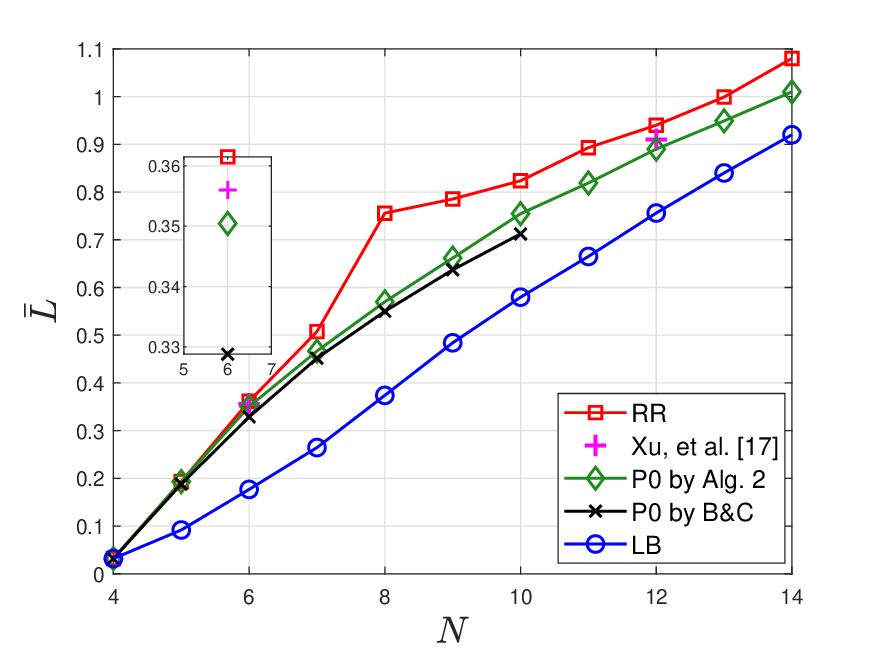}}
\caption{Expected coded shuffling load $\bar{L}$ vs. number of files  $N$  ($K=4$, mapping load $[M_1,M_2,M_3,M_4]=[3 , 4, 4,5]$, (normalized) reducing load $[W_1,W_2,W_3,W_4]=[1/8, 1/4, 1/4, 3/8]$, and Zipf parameter $\theta=0.56$).}
\label{fig:data1}
  \centerline{\includegraphics[width=1\linewidth]{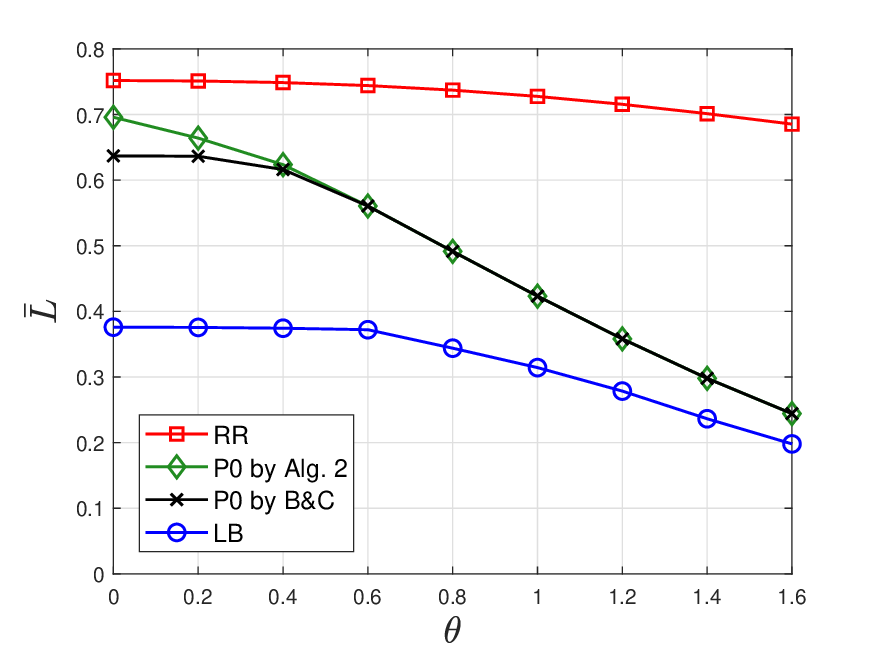}}
\caption{Expected shuffling load $\bar L$ vs. Zipf parameter $\theta$ ($N=8$, $K=4$,  mapping loads $[M_1,M_2,M_3,M_4]=[3 , 4, 4,5]$, and (normalized) reducing load $[W_1,W_2,W_3,W_4]=[1/8, 1/4, 1/4, 3/8]$). }
\label{fig:data_theta_1}
\end{figure}

\begin{table*}[t]
\centering
\caption{Computation time ( sec.)  of the two approaches for optimizing the heterogeneous CDC (the same setup as that in Fig.~\ref{fig:data1}). }
\vspace{-0.9em}
\label{table1}
\begin{tabular}{|p{15.8em}<{\centering}|p{2.5em}<{\centering}|p{2.5em}<{\centering}|
p{2.5em}<{\centering}|p{2.5em}<{\centering}|p{2.5em}<{\centering}|p{2.8em}<{\centering}
|p{2.8em}<{\centering}|p{2.5em}<{\centering}|p{2.5em}<{\centering}|p{2.5em}<{\centering}
|p{2.5em}<{\centering}|}
\hline
$N$  &4  & 5 &6& 7&8 & 9 &10& 11 & 12 & 13&14 \\ \hline
{\bf P0} with branch-and-cut &3 &9 &298  & 9,245&71,321  &162,212  &349,222  & N/A & N/A   & N/A  & N/A   \\ \hline
Algorithm~\ref{alg:p0} (two-file-group-based)  &2             & 3& 5&11 &17 & 30  & 45  & 61   & 103   & 182&279   \\ \hline
\end{tabular}
\end{table*}

We now evaluate the performance of the proposed heterogeneous CDC and C-CDC schemes. We generate the popularity distribution of files $\pbf$ using  Zipf distribution with $p_n={n^{-\theta}}/{\sum_{i=1}^{N}i^{-\theta}}$, where $\theta$ is the Zipf parameter. We assume that these files are accessed  by each job independently. Since the set of files required by each job is non-empty ($\Dc\ne\emptyset$), the probability of file subset $\Dc\subseteq\Nc$ being accessed by a job  is
$$p_{\Dc}=\frac{\prod_{n\in\Dc}p_{n}\prod_{n\notin\Dc}(1-p_{n})}{1-\prod_{n\in\Nc}(1-p_n)}.$$
The mapping load $M_k$ and the (normalized) reducing load $W_k$ for each worker $k$ will be set in each of our experiments below. 
\subsection{Heterogeneous CDC}\label{sec:simu1}
We first evaluate the performance of the proposed heterogeneous CDC scheme. For the proposed CDC scheme, we evaluate two methods we have proposed for optimizing the file placement and shuffling strategies: i) {\bf P0 by B\&C:} Solving {\bf P0} directly via the  branch-and-cut method by the standard optimization solver  MOSEK~\cite{Weise:MOSEK21};  ii) {\bf P0 by Alg. 2:} Algorithm~\ref{alg:p0} with the proposed two-file-group-based approach. 
For comparison, we also consider two existing file placement strategies for the proposed heterogeneous CDC:
\begin{enumerate}
\item 
{\bf Xu, et al~\cite{Xu&Shao:TCOM21}:} the file placement strategy proposed in~\cite{Xu&Shao:TCOM21}. Note that this strategy can only be used for certain values of $N$.

\item
{\bf Round robin (RR):} the round-robin file placement strategy commonly used by uncoded distributed computing~\cite{Anan2011:EuroSys}.
\end{enumerate}
Given the above file placement strategies, $\{t_{n,\Sc},n\in\Nc,\Sc\subseteq\Kc\}$ is known. In this case,  {\bf P0} is reduced to a nested coded delivery strategy optimization problem to solve, which is an LP problem. Besides the above two schemes, we also consider the following benchmark:
\begin{enumerate}[resume]
\item 
{\bf Lower bound (LB):} by relaxing the binary constraint \eqref{eq:con1} for $t_{n,\Sc}$ in {\bf P0} to continuous variable $t_{n,\Sc} \in [0,1]$, $\forall n,\Sc$, we relax {\bf P0} to an LP problem.  This relaxed problem provides a lower bound for {\bf P0} and is served as a benchmark for our proposed schemes. {Note that in general, unless the optimal solution to the relaxed LP problem happens to be a binary solution, which is rare, there will be a gap between the relaxed problem and the original {\bf P0}. The number of $t_{n,S}$'s is $(2^K-1)N$. When this number becomes large, it may lead to more relaxation and a
more noticeable gap in certain system settings.}
\end{enumerate}

We consider a network of $K=4$ workers with nonuniform mapping loads
$[M_1,M_2,M_3,M_4]=[3 , 4, 4,5]$ and (normalized) reducing load $[W_1,W_2,W_3,W_4]=[1/8, 1/4, 1/4, 3/8]$. Fig.~\ref{fig:data1} shows the expected shuffling load $\bar L$ vs. the  number of files $N$ for  $\theta=0.56$. We see that our proposed CDC scheme optimized via {\bf P0} achieves the lowest $\bar L$ among all the schemes. At the same time,    the performance of Algorithm~\ref{alg:p0} is nearly identical to the branch-and-cut method, with only a small gap observed at $N=6,10$. This shows the effectiveness of the proposed low-complexity two-file-group-based approach. Note that the file placement strategy in~\cite{Xu&Shao:TCOM21} is only applicable  to $N=6$ and $12$. Thus, its performance is only shown for these two values of $N$, and the corresponding shuffling load is higher than our proposed schemes. Round robin performs the worst among  all the schemes considered, which is as expected. The performance gap between  {\bf P0} and the lower bound remains approximately constant for $N>6$ and is slightly reduced as $N$ becomes large.

For the system setup in Fig.~\ref{fig:data1}, we compare the computation
time of the two proposed optimization methods for the heterogeneous CDC, \ie solving {\bf P0} via branch-and-cut and Algorithm~\ref{alg:p0} with the two-file-group-based approach. The computation time for different values of $N$ is shown in Table~\ref{table1}. The simulation has been conducted using MATLAB R2021a on a Windows x64 machine of Intel i5 CPU
with 3.5 GHz and 32 GB RAM. We see that the computational complexity of the
branch-and-cut method increases significantly  with $N$, rendering it impractical for a moderately large value of $N$. In contrast, Algorithm~\ref{alg:p0} with the
 two-file-group-based approach has a very low computational complexity
that only increases mildly with $N$.

To study the effect of different levels of file popularity distributions on the performance, in Fig.~\ref{fig:data_theta_1}, we plot $\bar L$ vs. Zipf parameter $\theta$ for $N=8$, where a lower value of $\theta$ means a more uniform file popularity distribution.~\footnote{Note that the file placement strategy from Xu, et al~\cite{Xu&Shao:TCOM21} cannot be applied to this case.} We see that $\bar L$ reduces as $\theta$ increases for all schemes. This is because that with a larger $\theta$, a smaller number of popular files are required by the jobs, which usually will result in a smaller number of IVs being shuffled.
We see that the performance gap between Algorithm~\ref{alg:p0} and branch-and-cut for solving {\bf P0} is only noticeable for $\theta\le 0.5$. This shows that the two-file-group-based placement performs especially well
for  the scenario with a more diverse file popularity distribution. As $\theta$ increases, the performance of our proposed two methods becomes much closer to the lower bound, indicating that the two-file-group-based file placement is close to optimal. To see why, intuitively, as $\theta$ becomes large, the popularity among files becomes more diverse. Only a few files are very popular, while the rest are unpopular. The two-file-group-based placement matches this underlying file popularity distribution and thus performs nearly optimal. In contrast, the gap between the round robin approach and our proposed two-file-group-based placement becomes significantly larger as $\theta$ increases. Since the round-robin
approach does not consider file popularity, its performance is not sensitive to different Zipf parameter $\theta$'s and always results in a much higher shuffling load.

\begin{table*}[t]
\centering
\caption{Computation time (sec.)  of the two approaches for optimizing the heterogeneous CDC (the same setup as that in Fig.~\ref{fig:c_cdc1}). }
\label{table:c_cdc1}
\vspace{-0.9em}
\begin{tabular}{|p{18.8em}<{\centering}|p{2.5em}<{\centering}|p{2.5em}<{\centering}|
p{2.5em}<{\centering}|p{2.5em}<{\centering}|p{2.5em}<{\centering}|p{2.8em}<{\centering}
|p{2.8em}<{\centering}|p{2.5em}<{\centering}|p{2.5em}<{\centering}|}
\hline
$K$  &4  & 5 &6& 7&8 & 9 &10& 11 & 12  \\ \hline
{\bf P3} with Branch-and-cut &6,528 &62,192 &231,588& 685,282&N/A  &N/A  &N/A  & N/A & N/A    \\ \hline
Algorithm~\ref{alg:p0} for {\bf P3} (two-file-group-based)  &2     & 5& 8&26 &46 & 74  & 138  & 222   & 312     \\ \hline
\end{tabular}
\end{table*}

\begin{table*}[t]
\centering
\caption{Computation time (sec.)  of the two approaches for optimizing the heterogeneous C-CDC (the same setup as that in Fig.~\ref{fig:c_cdc3}). }
\vspace{-0.9em}
\label{table2}
\begin{tabular}{|p{18.8em}<{\centering}|p{2.5em}<{\centering}|p{2.5em}<{\centering}|
p{2.5em}<{\centering}|p{2.5em}<{\centering}|p{2.5em}<{\centering}|p{2.8em}<{\centering}
|p{2.8em}<{\centering}|p{2.5em}<{\centering}|p{2.5em}<{\centering}|}
\hline
$N$  &5  & 6 &7& 8&9 & 10   \\ \hline
{\bf P3} with Branch-and-cut &826 &9,112 &61,588& 285,282&N/A  &N/A      \\ \hline
Algorithm~\ref{alg:p0} for {\bf P3} (two-file-group-based)  &3     & 7& 13&20 &31 & 49    \\ \hline
\end{tabular}
\end{table*}

\subsection{Heterogeneous C-CDC}
\begin{figure}
  \centerline{\includegraphics[width=1\linewidth]{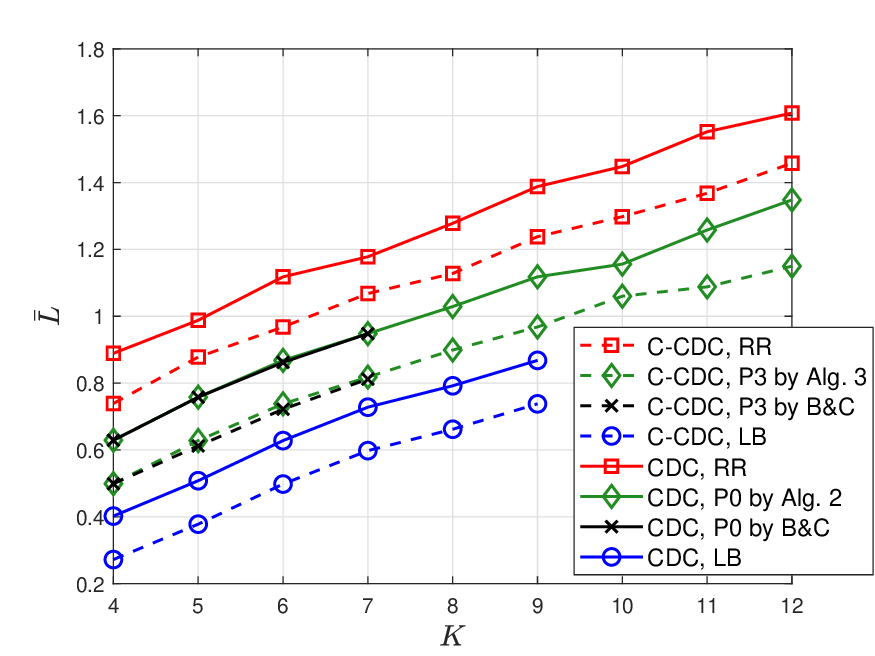}}
\caption{Expected shuffling load $\bar{L}$ vs. number of workers $K$ ($N=7$, Zipf parameter $\theta=0.56$. $M_k=3$, $k\le\lfloor\frac{K}{2}\rfloor$; $M_k=4$, $k\ge\lfloor\frac{K}{2}\rfloor+1$. $W_k=1/K$, $k\in\Kc$).}
\label{fig:c_cdc1}
\end{figure}

We now evaluate the performance of the proposed heterogeneous C-CDC. {We optimize the C-CDC in {\bf P3} by two methods: i) {\bf P3 by B\&C:} the branch-and-cut method (using MOSEK~\cite{Weise:MOSEK21}) and ii) {\bf P3 by Alg. 3:} Algorithm~\ref{alg:p1} (the two-file-group-based approach) applied to {\bf P3}. For comparison, we again consider iii) {\bf RR:} round-robin file placement;  iv) {\bf Lower bound (LB):} the relaxed problem of {\bf P3}, where the binary constraint \eqref{eq:con1} for $t_{n,\Sc}$ is relaxed to continuous variable $t_{n,\Sc} \in [0,1]$, $\forall n,\Sc$, which leads to an LP problem.} The above four schemes are also applied to the heterogeneous CDC for comparison. 

In Fig.~\ref{fig:c_cdc1}, we plot the expected shuffling load $\bar L$ vs. the number of workers $K$. We set $N=7$
and Zipf parameter $\theta=0.56$. For each $K$ value, we set the workers' mapping loads  by $M_k=3$ for $\forall k\le\lfloor\frac{K}{2}\rfloor$ and $M_k=4$ for $\forall k\ge\lfloor\frac{K}{2}\rfloor+1$, and set the (normalized) reducing load by $W_k=1/K$ for $k\in\Kc$. We only consider the branch-and-cut method for $K\le7$, for both {\bf P0} and {\bf P3}, due to its high computational complexity. We see that the C-CDC results in a lower shuffling load than the CDC, with the gap observed for all four different schemes.   This  reduction comes from the use of IV aggregation in the C-CDC. For the C-CDC optimized via {\bf P3}, Algorithm~\ref{alg:p0} performs very closely to the branch-and-cut method. This demonstrates the effectiveness of the two-file-group-based approach, which can achieve a good performance with much lower computational complexity than the branch-and-cut method. As expected, we see Algorithm~\ref{alg:p0} and the branch-and-cut method are outperforms the round robin placement, with an approximately constant gap for all values of $K$ considered. For the setup in Fig.~\ref{fig:c_cdc1}, the computation time for {\bf P3} by  Algorithm~\ref{alg:p0}  and the branch-and-cut method over $K$ are compared in Table~\ref{table:c_cdc1}. We again see that Algorithm~\ref{alg:p0} has  significantly lower computational complexity than the branch-and-cut method and is scalable over $K$, with a slower rate of complexity increment. 

\begin{figure}[t]
  \centering
  \centerline{\includegraphics[width=.967\linewidth]{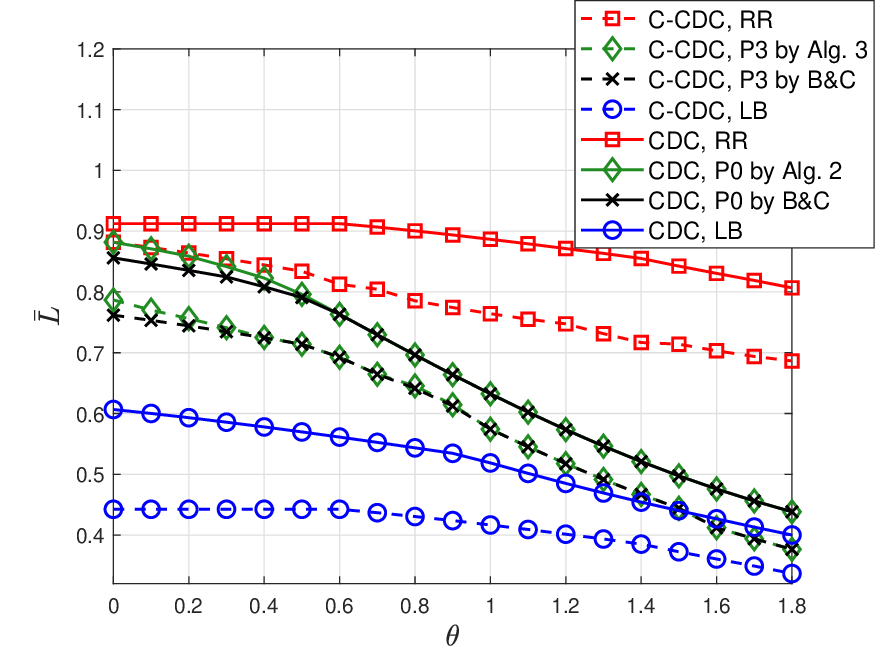}}
\caption{Expected coded shuffling load $\bar{L}$ vs. Zipf parameter $\theta$ ($N=8$, $K=6$, mapping load $[M_1,M_2,M_3,M_4,M_5]=[3,3,4,4,5,5]$, (normalized) reducing load $[W_1,W_2,W_3,W_4,W_5]=[1/12,1/12,1/6,1/6,1/4,1/4]$).}
\label{fig:c_cdc2}
 \end{figure}
 \begin{figure}
  \centerline{\includegraphics[width=1\linewidth]{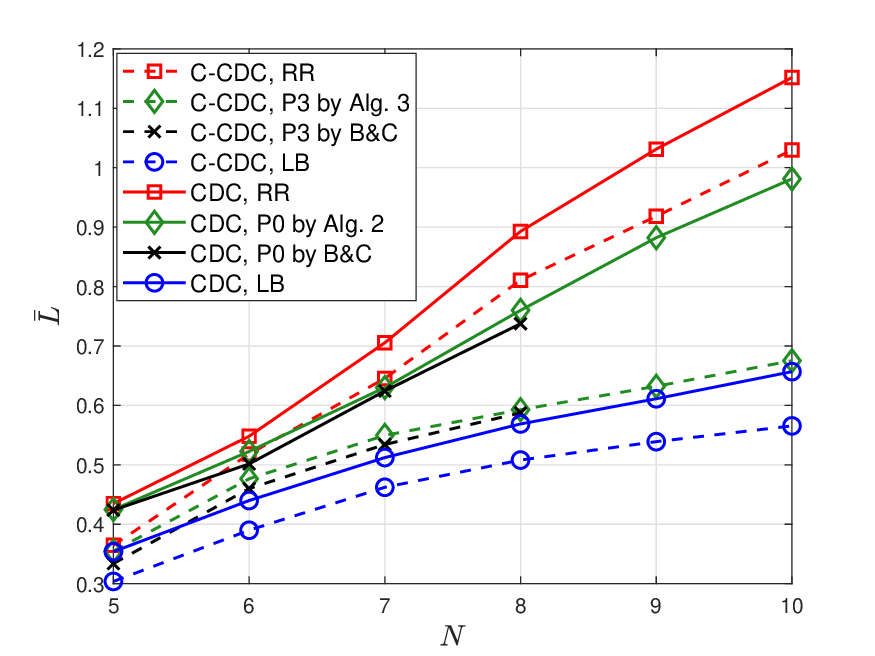}}
\caption{Expected coded shuffling load $\bar{L}$ vs. number of fils $N$ (Zipf parameter $\theta=0.8$, $K=5$, mapping load $[M_1,M_2,M_3,M_4,M_5]=[3,3,4,5,5]$, (normalized) reducing load $[W_1,W_2,W_3,W_4,W_5]=[1/10,1/10,1/5,1/5,2/5]$).}
\label{fig:c_cdc3}
\end{figure}

In Fig.~\ref{fig:c_cdc2}, we plot $\bar L$ vs. Zipf parameter $\theta$ for $N=8$ files. We consider $K=6$ workers with mapping loads $[M_1,M_2,M_3,M_4,M_5]=[3,3,4,4,5,5]$ and (normalized) reducing loads $[W_1,W_2,W_3,W_4,W_5]=[1/12,1/12,1/6,1/6,1/4,1/4]$. We again see that the C-CDC results in lower $\bar L$'s as compared with the CDC for all $\theta$'s considered. Between the two proposed methods for optimizing the C-CDC, we again see $\bar L$ by Algorithm~\ref{alg:p0} is only slightly higher than that of the branch-and-cut method for a small $\theta\in[0,0.6]$, and the two methods overlaps for $\theta>0.6$. This indicates the effectiveness of the two-file-group-based cache placement for a wide range of file popularity distribution. In general, the round-robin placement is less sensitive to the change of file popularity distribution, and its shuffling load does not reduce significantly as $\theta$ increases (more diverse file popularity distribution). 

In Fig.~\ref{fig:c_cdc3}, we plot $\bar L$ vs. the number of files $N$. We set the Zipf parameter $\theta=0.8$ and $K=5$. We also set the workers' mapping load $[M_1,M_2,M_3,M_4,M_5]=[3,3,4,5,5]$ and (normalized) reducing load vector $[W_1,W_2,W_3,W_4,W_5]=[1/10,1/10,1/5,1/5,2/5]$. We only consider the branch-and-cut method for $N=5,6,7,8$, due to its high computational complexity. We observe that the performance of  Algorithm~\ref{alg:p0} and the branch-and-cut method are very close for different values of $N$. For the optimized C-CDC using these two methods, we see it provides a considerable shuffling load reduction over the optimized CDC, especially for larger $N$. The optimized C-CDC follows the lower bound with a relatively small constant gap, in contrast to the optimized CDC, which has a much bigger gap to its lower bound. In Table~\ref{table2}, we compare the computation time for {\bf P3} by Algorithm~\ref{alg:p0}   and the branch-and-cut method over $N$ for the setup in Fig.~\ref{fig:c_cdc3}. We again see the computational complexity of Algorithm~\ref{alg:p0}    is significantly lower and only increases slowly as $N$ increases.

\begin{figure}[t]
  \centering
  {\includegraphics[width=1\linewidth]{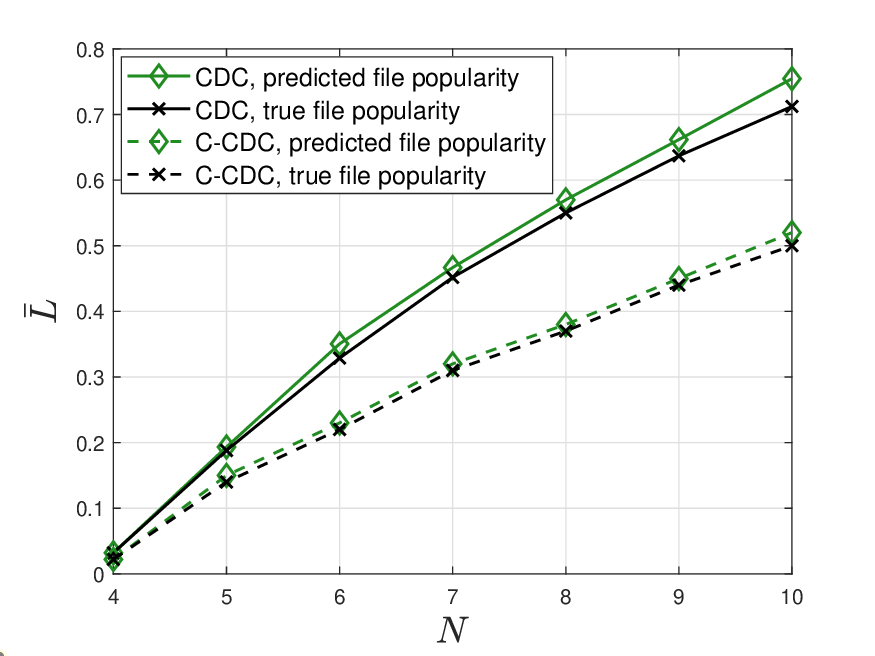}}
\caption{{Expected coded shuffling load $\bar{L}$ vs. number of files  $N$  ($K=4$, mapping load $[M_1,M_2,M_3,M_4]=[3 , 4, 4,5]$, (normalized) reducing load $[W_1,W_2,W_3,W_4]=[1/8, 1/4, 1/4, 3/8]$.}}
\label{fig:inaccu_popu}
\end{figure}

{\emph{Impact of imperfect knowledge of file popularity distribution:} Finally, we evaluate the performance of the proposed CDC and C-CDC assuming they are optimized based on an inaccurate popularity distribution. This is motivated by the fact that, in practice, the file popularity tends to be time-varying, and the accurate popularity distribution may not always be available. We assume that the Zipf parameter of the file popularity distribution $\pbf$ is predicted as $\hat\theta=0.7$, while the its true value is $\theta=0.56$. We solve {\bf P0} by Algorithm 2 based on $\hat\theta=0.7$ to obtain the  file placement strategy $\{t_{n,\Sc}\}$ and nested coded shuffling strategy $\{L_{k,\Sc}^{\Dc}, r_{k,\Sc}^{\Sc_{\!-\!i}}(\Dc)\}$. Then, we evaluate the average shuffling load  $\bar L$ in~\eqref{equ:avg_L} under $\theta=0.56$ using the obtained strategy solution. A similar evaluation is conducted for the C-CDC for {\bf P3}
using Algorithm 3. We set $K=4$ workers with nonuniform mapping loads
$[M_1,M_2,M_3,M_4]=[3 , 4, 4,5]$ and (normalized) reducing loads $[W_1,W_2,W_3,W_4]=[1/8, 1/4, 1/4, 3/8]$.  As shown in Fig.~\ref{fig:inaccu_popu}, for both CDC and C-CDC, using predicted   $\hat\theta=0.7$,  the resulting average shuffling load $\bar L$ is slightly higher than that under the true $\theta=0.56$.  For the CDC, the gap is slightly larger as $N$ becomes larger, while for the C-CDC, the gap remains small for all the values of $N$ considered. This may be because more IVs need to be shuffled under the
CDC, which creates more shuffling load under popularity
mismatch. In general, the gap due to inaccurate file popularity distribution prediction is very small, showing a robust performance under a moderate inaccurate file popularity distribution.}

\section{Conclusion}\label{sec:conclusion}
In this paper, we have investigated the design of heterogeneous CDC, where the popularity of files to be accessed by the jobs is  nonuniform. Different from the existing works, we have proposed a mapping strategy that can handle an arbitrary number of files. Furthermore, we provide a nested coded shuffling strategy to efficiently handle the files with nonuniform popularity and maximize the coded multicasting opportunity. We further optimize the proposed CDC scheme by formulating a joint optimization problem of the file placement and nested shuffling parameters.  Since solving this joint optimization problem via the conventional branch-and-cut method incurs high computational complexity, we have proposed a low-complexity two-file-group-based file placement approach to find a solution. Numerical studies show that the two-file-group-based approach provides a performance that is very close to the branch-and-cut method but with significantly lower computational complexity, which is scalable over both $N$ and $K$. Furthermore, our optimized CDC scheme outperforms other alternative methods.

 For the computing jobs with aggregate target functions, we have proposed a heterogeneous C-CDC scheme to further improve the shuffling efficiency.  The C-CDC uses a local IV aggregation technique to compress the IVs before coded shuffling, such that the shuffling load can be further reduced. We again optimize the proposed C-CDC over the file placement and nested shuffling parameters and explore the two-file-group-based low-complexity approach to solve the problem.
Numerical results demonstrate that a considerable shuffling load reduction can be achieved by the C-CDC over the CDC, and the two-file-group-based file placement approach maintains a good performance for the C-CDC as well.
\balance
\bibliographystyle{IEEEtran}
\bibliography{Yong}
\vspace{-0em}
\begin{IEEEbiographynophoto}
{Yong Deng} (Member, IEEE) received the Ph.D. degree in electrical and computer engineering from Ontario Tech University, Canada, in 2021. Before that, he received the M.S. degree in Computer Science from Zhejiang Gongshang University, China, in 2016. From 2016 to 2017, he was a Visiting Researcher with Faculty of Business and Information Technology at Ontario Tech University. From 2021 to 2022, he was a Post-Doctoral fellow with the Edward S. Rogers Sr. Department of Electrical and Computer engineering, University of Toronto, Canada. From 2022 to 2023, he was an Assistant Professor (tenure-track) with the Department of Computer Science, St. Francis Xavier University, Canada. He is currently an Assistant Professor (tenure-track) with the Department of Software Engineering, Lakehead University, Canada.  His research interests include optimization, coded caching, coded distributed computing, mechanism design in wireless networks and network security.
\end{IEEEbiographynophoto} 
 
 \begin{IEEEbiographynophoto}
 {Min Dong} (Fellow, IEEE) received the  B.Eng. degree from Tsinghua University, Beijing, China, in 1998, and the  Ph.D. degree in electrical and computer engineering with a minor in applied  mathematics from Cornell University, Ithaca, NY, in 2004. From 2004 to 2008, she was with Qualcomm Research, Qualcomm Inc., San Diego, CA. Since 2008,  she has been with Ontario Tech University, where she is currently a Professor  with the Department of Electrical, Computer and Software Engineering. Her  research interests include wireless communications, statistical signal processing, learning techniques, optimization and control applications in cyber-physical  systems. 

Dr. Dong received the Early Researcher Award from the Ontario Ministry of Research and Innovation in 2012, the Best Paper Award at IEEE ICCC in 2012, and the 2004 IEEE Signal Processing Society Best Paper Award. She is a co-author of the Best Student Paper at IEEE SPAWC 2021 and the Best Student Paper of Signal Processing for Communications and Networking at IEEE ICASSP 2016. She is a Senior Area Editor for IEEE TRANSACTIONS ON SIGNAL PROCESSING and an Associate Editor for the IEEE OPEN JOURNAL of SIGNAL PROCESSING. She served as an Editor for IEEE TRANSACTIONS ON WIRELESS COMMUNICATIONS (2018-2023) and as an Associate Editor for the IEEE TRANSACTIONS ON SIGNAL PROCESSING (2010-2014) and the IEEE SIGNAL PROCESSING LETTERS (2009-2013). She was also on the Steering Committee of the IEEE TRANSACTIONS ON MOBILE COMPUTING (2019-2021). She was an elected member of the Signal Processing for Communications and Networking (SP-COM) Technical Committee of IEEE Signal Processing Society (2013-2018). She was the lead Co-Chair of the Communications and Networks to Enable the Smart Grid Symposium at the IEEE International Conference on Smart Grid Communications in 2014.  
\end{IEEEbiographynophoto}
\end{document}